\documentclass[11pt,aps,showpacs,nofootinbib,prd,aps,epsf,floats,axodraw,amsmath,amssymb,amsfonts]{revtex4}
\usepackage{amsmath, amssymb}
\usepackage{axodraw}
\newcommand{\mathsym}[1]{{}}

\usepackage{graphicx}
\usepackage{amsmath}
\usepackage{amssymb}
\usepackage{bm}
\newcommand{\be}{\begin{equation}}
\newcommand{\ee}{\end{equation}}
\newcommand{\bea}{\begin{eqnarray}}
\newcommand{\eea}{\end{eqnarray}}

\newcommand{\rem}[1]{}
\newsavebox{\PSLASH}
 \sbox{\PSLASH}{$p$\hspace{-1.8mm}/}
 
\renewcommand{\theequation}{\thesection.\arabic{equation}}
\newcounter{saveeqn}
\newcommand{\add}{\addtocounter{equation}{1}}
\newcommand{\alpheqn}{\setcounter{saveeqn}{\value{equation}}%
\setcounter{equation}{0}%
\renewcommand{\theequation}{\mbox{\thesection.\arabic{saveeqn}{\alph{equation}}}}}
\newcommand{\reseteqn}{\setcounter{equation}{\value{saveeqn}}%
\renewcommand{\theequation}{\thesection.\arabic{equation}}}

 \newsavebox{\notrightarrow}
 \sbox{\notrightarrow}{$\to$\hspace{-4mm}/}
 
 \newsavebox{\PARTIALSLASH}
 \sbox{\PARTIALSLASH}{$\partial$\hspace{-1.6mm}/}
 
 \newsavebox{\ASLASH}
 \sbox{\ASLASH}{$A$\hspace{-2.1mm}/}
 
 \newsavebox{\KSLASH}
 \sbox{\KSLASH}{$k$\hspace{-1.8mm}/}
 
 \newsavebox{\LSLASH}
 \sbox{\LSLASH}{$\ell$\hspace{-1.8mm}/}
 
 \newsavebox{\QSLASH}
 \sbox{\QSLASH}{$q$\hspace{-1.8mm}/}
 
 \newsavebox{\DSLASH}
 \sbox{\DSLASH}{$D$\hspace{-2.2mm}/}
 
 \newsavebox{\DbfSLASH}
 \sbox{\DbfSLASH}{${\mathbf D}$\hspace{-2.8mm}/}
 
 \newsavebox{\DELVECRIGHT}
 \sbox{\DELVECRIGHT}{$\stackrel{\rightarrow}{\partial}$}
 
 \newcommand{\blue}{\IfColor{\textCadetBlue}{}}
\newcommand{\black}{\IfColor{\textBlack}{}}
\newcommand{\red}{\IfColor{\textRed}{}}
\newcommand{\green}{\IfColor{\textOliveGreen}{}}
\newcommand{\lila}{\IfColor{\textRedViolet}{}}

\begin{document}
\title{Translational-invariant noncommutative gauge theory}
\author{F. Ardalan$^{a,b}$}\email{ardalan@ipm.ir}
\author{N. Sadooghi$^{a}$}\email{sadooghi@physics.sharif.ir}
\affiliation{$^a$Department of Physics, Sharif University of
Technology, P.O. Box 11155-9161, Tehran-Iran\\
$^{b}$Institute for Studies in Theoretical Physics and Mathematics (IPM)\\
School of Physics, P.O. Box 19395-5531, Tehran-Iran}

\begin{abstract}
A generalized translational invariant noncommutative field theory is
analyzed in detail, and a complete description of translational
invariant noncommutative structures is worked out.  The relevant
gauge theory is described, and the planar and nonplanar axial
anomalies are obtained.\\
Keywords: Non-Commutative Geometry, Space-time Symmetries\\
E-print (arXiv) Number: 1008.5064 [hep-th]\\
\end{abstract}
\pacs{11.10.Nx, 11.10.Lm, 11.15.Bt, 11.15.Kc} \maketitle
\section{Introduction}\label{introduction}
\par\noindent
Noncommutative geometry \cite{connes} has a long history. The advent
of the flurry of activity in this field related to physics was the
discovery of noncommutativity in string theory \cite{string-nc}.
Subsequently, noncommutative field theories, which appear in a
decoupling limit of string theories, have been the focus of
extensive research. Noncommutative geometry is most elegantly
described in the context of noncommutative algebra \cite{connes},
and in particular noncommutative algebra of functions, which is an
ingenious generalization of the commutative $C^{*}$-algebra of
ordinary function as the Gelfand-Naimark dual of ordinary
commutative geometry. Thus, general noncommutative star-algebras of
functions are primary objects in noncommutative geometry and in
particular in the field theories based on such geometries. The first
example of the noncommutative geometry arising from string theory
was constructed on the basis of the canonical commutation relation
\begin{eqnarray}\label{I1}
[x_{\mu},x_{\nu}]_{\star}=i\theta_{\mu\nu},
\end{eqnarray}
where, in the simplest case, $\theta_{\mu\nu}$ is a constant real
antisymmetric matrix, and the Moyal product of functions derived
from (\ref{I1}) is \cite{moyal}
\begin{eqnarray}\label{I2}
\left(f\star
g\right)(x)=f(x)\exp\left(\frac{i}{2}~\theta_{\mu\nu}\overleftarrow{\partial}^{\mu}\overrightarrow{\partial}^{\nu}\right)g(x).
\end{eqnarray}
This noncommutative product of functions has the additional
properties of star-algebra
\begin{eqnarray}\label{I3}
(f\star g)^{*}=g^{*}\star f^{*}, \qquad\mbox{and}\qquad f^{**}=f.
\end{eqnarray}
In addition the function space is usually a unital algebra,
\begin{eqnarray}\label{I4}
f\star 1=1\star f=f.
\end{eqnarray}
The three properties (\ref{I1})-(\ref{I3}) are essential properties
of a noncommutative star-algebra, which when added with a norm
define a Banach algebra; a cornerstone of noncommutative geometry.
\par
At this stage, let us notice that the Moyal star-product (\ref{I2})
is not the unique choice compatible with (\ref{I1}). A second
alternative to quantize the classical Poisson structure is the
Wick-Voros product \cite{wick-voros}. In \cite{galluccio2008}, a
noncommutative $\lambda\varphi^{4}$ theory is formulated using these
two products, and the differences between them are studied. It turns
out that whereas the Lagrangian densities of these two apparently
different theories, and consequently their tree level vertices and
propagators as well as their one-loop Green's functions are
different, they have the same $S$-matrix element. They are therefore
``physically'' equivalent. Moreover, it is shown that since both
products are the realization of the same canonical commutation
relation (\ref{I1}), the one-loop Feynman integrals arising from
these two formulations have the same ultraviolet (UV) behavior. This
is why the UV/IR mixing \cite{minwalla1999}, appearing in the more
elaborated Moyal formulation cannot be cured in the Wick-Voros
formulation. To have a satisfactory interpretation of these
remarkable results, the authors in \cite{galluccio2008} use a
symmetry argument. They relate the equivalence between Moyal and
Wick-Voros formulations at the level of $S$-matrix elements to the
invariance of the physical observables, in general, and the
$S$-matrix elements, in particular, under the
Poincar$\acute{\mbox{e}}$ transformation. To prove this equivalence,
they use instead of (\ref{I1}), which is not invariant under the
``ordinary'' Poincar$\acute{\mbox{e}}$ transformation, a twisted
theory \cite{20-22}, which is formulated so that it is invariant
under a certain twisted Poincar$\acute{\mbox{e}}$ symmetry
\cite{23-25}. The latter is based on a deformed
Poincar$\acute{\mbox{e}}$ Lie algebra, that builds a noncommutative,
noncocommutative Hopf algebra. Using a consistent twisting
procedure, where the field operators (oscillators) of the theory are
also deformed and new commutation relations between creation and
annahilation operators are defined, they finally recalculate the
twisted $S$-matrix element, which is shown to have the same
expression as in the Moyal and Wick-Voros case.  Different twisted
formulations of quantum field theories are discussed in
\cite{26-35}. The deformation of field oscillators in a more general
framework of braided algebras is discussed in \cite{lukierski}.
\par
In this paper, we will use a third, more general product than Moyal
and/or Wick-Voros products. It is based on the crucial requirement,
that ensures the existence of energy-momentum conservation in the
usual sense, and that is the property of translation invariance
\begin{eqnarray}\label{I5}
{\cal{T}}_{a}(f)\star{\cal{T}}_{a}(g)={\cal{T}}_{a}\left(f\star
g\right),
\end{eqnarray}
where
\begin{eqnarray}\label{I6}
{\cal{T}}_{a}(f)(x)\equiv f(x+a).
\end{eqnarray}
General translational invariant associative star-product was
originally introduced in \cite{lizzi2009, translational invariant}
in terms of a certain function $\alpha(p,q)$, which is constrained
mainly by the associativity requirement of the product. Constructing
a simple scalar field theory using this new product, it was further
shown that the nonplanar Feynman integrals of the theory are mainly
modified by a combination of $\alpha(p,q)$ that reproduces, in
particular, the same antisymmetric phase factor that appears in the
Feynman integrals of noncommutative gauge theories constructed by
the ordinary Moyal product. This phase factor, given by the
commutator of coordinates (\ref{I1}) is responsible for the famous
UV and IR connection of noncommutative field theory
\cite{minwalla1999}. In the present work, we study these theories
further. Apart from generalizing the results arising from
noncommutative translational-invariant bosonic formulation to $U(1)$
gauge theory, the goal is to present the general structure of the
characteristic function $\alpha(p,q)$ in order to understand the
relation between the novel translational-invariant noncommutative
product with the other two, Moyal and Wick-Voros, products. This is
something which is not completely discussed in \cite{lizzi2009}. In
the light of this comparison, and following the line of arguments in
\cite{galluccio2008} (see also our descriptions above), the fact
that even the translational-invariant formulation is not able to
cure the UV/IR mixing of noncommutative field theories will be
clarified.
\par
The paper is organized as follows: In Sec. II, after determining the
general structure of the new translational invariant noncommutative
product in terms of the real and the imaginary part of
$\alpha(p,q)$, we perform a complete specification of the product
structure for two noncommutative directions, and determine in this
way the general solution of the cocycle relation which is, as before
mentioned, the main restriction on $\alpha(p,q)$. The main result of
this section is Eq. (\ref{Sx-54}), which states that the
noncommutative structure function $\alpha(p,q)$ is the sum of a
quadratic term $\omega(p,q)$ which enters in the loop diagrams and
an arbitrary complex function $\eta(p)$, with real part even and
imaginary part odd parity, that does not appear in the loop
integrations. In particular, the real even part of $\eta(p)$ seems
to be the generalization of the phase factor appearing in the
three-level propagator of noncommutative field theory formulated
with Wick-Voros product. In Secs. III and IV, we construct the
noncommutative gauge theory and its one-loop Feynman diagrams. Our
goal is to study the effect of the elements of the characteristic
function $\alpha(p,q)$ on the divergence properties of the Feynman
integrals. We will determine the loop integrations of one-loop
corrections to the fermion and photon propagators and vertex
function and present general arguments to show that loop
integrations in any order of perturbation theory involve only
$\omega(p,q)$ and not $\eta(p)$. We conclude that only $\omega(p,q)$
is responsible for the well-known noncommutative UV/IR mixing
\cite{minwalla1999}, and $\eta(p)$ does not play any r\^{o}le in the
divergence properties of Feynman integrals. Section V consists of a
study of axial anomalies of these gauge theories. Quantum anomalies
of the ordinary Moyal noncommutative gauge theories are studied
intensively in \cite{sadooghi2000,sadooghi2001,nc-anomaly}, where it
is shown that they consist of a planar (covariant) as well as a
nonplanar (invariant) anomaly. As in the ordinary Moyal
noncommutative gauge theory, we will show in Sec. V that whereas the
planar anomaly is a noncommutative generalization of the well-known
Adler-Bell-Jackiw axial anomaly, the nonplanar axial anomaly
consists of a generalized star-product \cite{garousi} now modified
with a phase factor consisting of a symmetric function in the
momenta. Section VI is devoted to concluding remarks.
\section{Translational invariant product}\label{review}
\subsection{General structure}
\par\noindent
\setcounter{equation}{0} In a translationally invariant
noncommutative product \cite{lizzi2009}, the kernel of the product
as defined by
\begin{eqnarray}\label{Sx-1}
f(x)\star g(x)\equiv \int
\frac{d^{d}p}{(2\pi)^{d}}~\frac{d^{d}q}{(2\pi)^{d}}~\frac{d^{d}r}{(2\pi)^{d}}~e^{-irx}~\tilde{f}(p)\tilde{g}(q)K\left(r,p,q\right),
\end{eqnarray}
has the following form
\begin{eqnarray}\label{Sx-2}
K\left(r,p,q\right)=e^{\hspace{0.05cm}\alpha(r,p)}\delta^{d}(p+q-r).
\end{eqnarray}
Translational invariance is defined by (\ref{I5})-(\ref{I6}). The
noncommutative star-product is therefore characterized by the
complex function $\alpha(p,q)$. The main restriction on
$\alpha(p,q)$ follows from associativity of the star-product
\begin{eqnarray}\label{Sx-5}
f\star(g\star h)=(f\star g)\star h,
\end{eqnarray}
which restricts the kernel function $K(r,p,q)$ by
\begin{eqnarray}\label{Sx-6}
\int d^{d}\ell~K\left(p,\ell,q\right)K\left(\ell,r,s\right)=\int
d^{d}\ell~K\left(p,r,\ell\right)K\left(\ell,s,q\right).
\end{eqnarray}
and imposes the associativity condition
\begin{eqnarray}\label{Sx-7}
\alpha(p,q)+\alpha(q,r)=\alpha(p,r)+\alpha(p-r,q-r),
\end{eqnarray}
on the characterizing function $\alpha(p,q)$. There is another
significant restriction on $\alpha(p,q)$ coming from the requirement
of the existence of conjugation on the function space (\ref{I3}),
which imposes the condition
\begin{eqnarray}\label{Sx-9}
\alpha(p,q)^{*}=\alpha(-p,q-p),
\end{eqnarray}
on the function $\alpha(p,q)$. Here, we will be requiring the
star-algebra to have the constant function $1$ as its identity [see
(\ref{I4})], resulting in
\begin{eqnarray}\label{Sx-11}
\alpha(p,p)=\alpha(p,0)=0.
\end{eqnarray}
The primary example of a noncommutative star-product is the Moyal
product defined by (\ref{I2}), which is equivalent to
\begin{eqnarray}\label{Sx-13}
\alpha(p,q)=-\frac{i}{2}\theta_{\mu\nu}p^{\mu}q^{\nu},
\end{eqnarray}
with $\theta_{\mu\nu}$ a constant antisymmetric matrix defined in
(\ref{I1}). In the rest of this section, we will find the most
general form of the complex function $\alpha(p,q)$, with the
additional assumption that $\alpha(p,q)$ can be expanded in a series
in the components of $p$ and $q$
\begin{eqnarray}\label{Sx-14}
\alpha(p,q)=\sum\limits_{n=0}^{\infty}\sum\limits_{\stackrel{\{i_{1},\cdots,
i_{n}\}}{\{j_{1},\cdots,j_{n}\}}}a_{i_{1},\cdots,i_{n};j_{1},\cdots
j_{n}}p_{1}^{i_{1}}\cdots p_{n}^{i_{n}}q_{1}^{j_{1}}\cdots
q_{n}^{j_{n}}.
\end{eqnarray}
In the expression (\ref{Sx-14}), there are no constant terms. This
is because of the unitality condition (\ref{Sx-11}). Moreover, each
term in the series contains at least one power of $p_{i}$ and one of
$q_{i}$.
\par
With these restrictions, the main constraint to be satisfied will be
the condition of associativity (\ref{Sx-7}), which we will proceed
to analyze. The first step in the analysis of associativity
condition is to separate the function $\alpha(p,q)$ in its real and
imaginary part
\begin{eqnarray}\label{Sx-15}
\alpha(p,q)=\alpha_{1}(p,q)+i\alpha_{2}(p,q),
\end{eqnarray}
where $\alpha_{1}(p,q)$ and $\alpha_{2}(p,q)$ are now real
functions.
\par\vspace{0.3cm}\noindent{\underline{\it{i) The real part of
$\alpha(p,q)$:}}
\par\vspace{0.3cm}\noindent
The condition of conjugation (\ref{Sx-9}) implies
\begin{eqnarray}\label{Sx-16}
\alpha_{1}(p,q)=\frac{1}{2}\big[\alpha(p,q)+\alpha(-p,q-p)\big].
\end{eqnarray}
Using the associativity condition (\ref{Sx-7}) by substituting $p\to
0, q\to q$, and $r\to p$, we get
\begin{eqnarray}\label{Sx-17}
\alpha(0,q)+\alpha(q,p)=\alpha(0,p)+\alpha(-p,q-p).
\end{eqnarray}
Using the associativity condition (\ref{Sx-7}) again, but this time
with the substitutions $p\to p, q\to q$, and $r\to p$, we have
\begin{eqnarray}\label{Sx-18}
\alpha(p,q)+\alpha(q,p)=\alpha(0,q-p),
\end{eqnarray}
where we have used condition for existing of identity (\ref{Sx-11}),
$\alpha(p,p)=0$. The net result is
\begin{eqnarray}\label{Sx-19}
\alpha_{1}(p,q)=\eta_{1}(q)-\eta_{1}(p)+\eta_{1}(p-q),
\end{eqnarray}
where $\eta_{1}(p)\equiv \frac{1}{2}~\alpha(0,p)$. We note that
$\eta_{1}(p)$ is real, by complex conjugation (\ref{Sx-9}), and an
even function of $p$,
\begin{eqnarray}\label{Sx-20}
\eta_{1}(-p)=\eta_{1}(p),
\end{eqnarray}
from the associativity condition (\ref{Sx-7}), again, this time with
the substitution $p\to r$, and $q\to 0$. It can be readily verified
that $\alpha_{1}(p,q)$ as given by (\ref{Sx-19}) satisfies the
associativity condition identically for arbitrary even function
$\eta_{1}(p)$. Moreover
\begin{eqnarray}\label{Sx-21}
\eta_{1}(0)=0,
\end{eqnarray}
by the existence of the unit of the algebra (\ref{Sx-11}). Therefore
the real part of $\alpha(p,q)$ is given by (\ref{Sx-19}) in terms of
an arbitrary real even function $\eta_{1}$ satisfying (\ref{Sx-21}).
Note that the function $\eta_{1}(p)$ plays the role of a weighting
function for the integral of the trace relation
\begin{eqnarray}\label{Sx-22}
\int d^{d}x~f(x)\star g(x)=\int d^{d}x~g(x)\star
f(x)=\int\frac{d^{d}p}{(2\pi)^{d}}~ e^{2\eta_{1}(p)}f(p)g(p).
\end{eqnarray}
For that reason, it effectively determines the function space on
which the star-algebra is built.
\par\vspace{0.3cm}\noindent{\underline{\it{ii) The imaginary part of
$\alpha(p,q)$:}}
\par\vspace{0.3cm}\noindent
The determination of the imaginary part of $\alpha(p,q)$ is more
involved. It was observed in \cite{lizzi2009} that only a certain
part of $\alpha_{2}(p,q)$ in (\ref{Sx-15}) defined by
\begin{eqnarray}\label{Sx-23}
-2i\omega(p,q)=\alpha(p+q,p)-\alpha(p+q,q),
\end{eqnarray}
appears in the loop integrals of the scalar $\lambda\varphi^{4}$
theory.\footnote{Note that in the definition of $\omega(p,q)$, there
is an additional $-2i$ in Ref. \cite{lizzi2009}.} We will show in
the subsequent section that this persists for the gauge theory also.
However, we will find that $\alpha_{2}(p,q)$, the imaginary part of
$\alpha(p,q)$, has an additional contribution that we call
$\xi(p,q)$,
\begin{eqnarray}\label{Sx-24}
\alpha_{2}(p,q)=\omega(p,q)+\xi(p,q).
\end{eqnarray}
We will proceed to determine the form of both $\omega(p,q)$ and
$\xi(p,q)$. In \cite{lizzi2009}, the form of $\omega(p,q)$ was
correctly identified; however, the arguments required are more rigor
and we will provide it.
\par
To begin with, it is straightforward to see that $\omega(p,q)$ is
real, as can be seen from the fact that the real part of
$\alpha(p+q,p)$ from (\ref{Sx-19}) is symmetrized in the exchange of
$p$ and $q$. It is also clear that $\omega(p,q)$ is antisymmetric in
$p\leftrightarrow q$. Thus:
\begin{eqnarray}
&\hspace{-1cm}\mbox{$\omega(p,q)$ is real},&\\ \label{Sx-25}
&\omega(p,q)=-\omega(q,p),\qquad\qquad&\mbox{$\omega(p,q)$ is
antisymmetric in $p$ and $q$},\label{Sx-26}
\end{eqnarray}
We can also show that $\omega(p,q)$ is an even function of $p$ and
$q$. First, we have
\begin{eqnarray*}
-2i\omega(-p,-q)=\alpha(-p-q,-p)-\alpha(-p-q,-q).
\end{eqnarray*}
But from associativity (\ref{Sx-7}), with the substitution $p\to 0,
q\to p$, and $r\to p+q$, we get
\begin{eqnarray}\label{Sx-27}
\alpha(-p-q,-q)=\alpha(p,p+q)+\alpha(0,p)-\alpha(0,p+q).
\end{eqnarray}
Then, using (\ref{Sx-18}), we get
\begin{eqnarray}\label{Sx-28}
\omega(-p,-q)=\omega(p,q).\qquad\qquad \mbox{$\omega(p,q)$ is odd in
$p$ and $q$.}
\end{eqnarray}
More significantly $\omega(p,q)$ satisfies the same associativity
relation (\ref{Sx-7}) as $\alpha(p,q)$:
\begin{eqnarray}\label{Sx-29}
\omega(p,q)+\omega(q,r)=\omega(p,r)+\omega(p-r,q-r).
\end{eqnarray}
To prove this, first we note that using the associativity relation
as
$$
\alpha(p+q,p)+\alpha(p,r)=\alpha(p+q,r)+\alpha(p+q-r,p-r)
$$
and substituting $r\to q$, we get
\begin{eqnarray}\label{Sx-30}
-2i\omega(p,q)=\alpha(p,p-q)-\alpha(p,q).
\end{eqnarray}
Then, proving associativity for $\omega(p,q)$ reduces to proving
that
$\alpha(p,p-q)+\alpha(q,q-r)-\alpha(p,p-r)\linebreak-\alpha(p-r,p-q)$
vanishes. Using now $-\alpha(p,p-r)=\alpha(p-r,p)-\alpha(0,r),$ we
get
$\alpha(p-r,p)\linebreak+\alpha(p,p-q)-\alpha(p-r,p-q)+\alpha(q,q-r)-\alpha(0,r)$,
which upon using associativity again becomes
$\alpha(q-r,q)+\alpha(q,q-r)-\alpha(0,r)$, which vanishes by
(\ref{Sx-18}). Thus we have proved (\ref{Sx-29}).
\par\noindent
Now clearly as $\alpha(p,q)$ satisfies associativity and also
$\alpha_{1}(p,q)$, we conclude that so does $\xi(p,q)$:
\begin{eqnarray}\label{Sx-31}
\xi(p,q)+\xi(q,r)=\xi(p,r)+\xi(p-r,q-r).
\end{eqnarray}
Antisymmetry of $\xi(p,q)$,
\begin{eqnarray}\label{Sx-32}
\xi(p,q)=-\xi(q,p),
\end{eqnarray}
follows from (\ref{Sx-18}) and antisymmetry of $\omega(p,q)$,
together with the relation
\begin{eqnarray*}
\alpha_{1}(p,q)+\alpha_{1}(q,p)=\alpha(0,p-q).
\end{eqnarray*}
The parity of $\xi(p,q)$,
\begin{eqnarray}\label{Sx-33}
\xi(-p,-q)=-\xi(p,q),
\end{eqnarray}
is obtained from the definition of $\omega(p,q)$ and $\xi(p,q)$,
which gives, using both (\ref{Sx-23}) and (\ref{Sx-30}),
\begin{eqnarray}\label{Sx-34}
\xi(p+q,p)=\xi(p+q,q).
\end{eqnarray}
\subsection{Determination of $\omega(p,q)$ and $\xi(p,q)$ for two
noncommutative dimensions}
\newcommand{\vp}{\vec{p}}
\newcommand{\vq}{\vec{q}}
\newcommand{\ver}{\vec{r}}
\par\noindent
From now on we assume that the noncommutativity occurs for only two
spatial coordinates $(x_{1},x_{2})$, and, assuming a series
expansion for $\omega(\vp,\vq)$ and $\xi(\vp,\vq)$, find their most
general form. Of the three basics conditions on
$\omega(\vec{p},\vec{q})$ and $\xi(\vec{p},\vec{q})$ the most
restrictive is the associativity condition (\ref{Sx-29}) and
(\ref{Sx-31}), which follows from associativity condition on
$\alpha(\vec{p},\vec{q})$, (\ref{Sx-7}), and the complex conjugation
condition (\ref{Sx-9}).\footnote{In two dimensions, $\vec{p}$
denotes $(p_1,p_2)$.} We will in fact see shortly that condition
(\ref{Sx-9}) is indeed incorporated in (\ref{Sx-29}) and
(\ref{Sx-31}). There remains to impose the condition of the unit of
star-algebra, (\ref{Sx-11}). Imposing this condition implies that in
each term in the series (\ref{Sx-14})
\begin{eqnarray}\label{Sx-35}
\sum\limits_{\stackrel{\{i_{1},i_{2},j_{1},j_{2}\}}{i_{1}+i_{2}+j_{1}+j_{2}=N}}a_{i_{1},i_{2};j_{1},j_{2}}p_{1}^{i_{1}}p_{2}^{i_{2}}q_{1}^{j_{1}}q_{2}^{j_{2}},
\end{eqnarray}
we have, $i_{1}+i_{2}>0$ as well as $j_{1}+j_{2}>0$. Note that we
have picked a particular term in the series (\ref{Sx-14}) of total
degree of $N$, as the associativity condition being linear operation
within terms of a fixed total degree $N$.\par The task on hand is
therefore to extract the restriction of associativity conditions
(\ref{Sx-29}) and (\ref{Sx-31}) on the coefficients
$a_{\vec{i},\vec{j}}$, where we denote $\vec{i}=(i_{1},i_{2})$ and
$\vec{j}=(j_{1},j_{2})$. We will show that for total degree $N$
even, appropriate for $\omega(p,q)$, the only solution of the
(\ref{Sx-29}) is
\begin{eqnarray}\label{Sx-36}
\omega(\vp,\vq)=\vp\wedge\vq\equiv \frac{\theta}{2}(p_1 q_2-p_2
q_1),
\end{eqnarray}
with $N=2$. Here, $\theta$ is a multiplicative constant. But, we see
that there are many solutions for $N$ odd, appropriate for
$\xi(\vp,\vq)$, and we will determine them. We will use the
polynomial (\ref{Sx-35}) for both $\omega(\vp,\vq)$ and
$\xi(\vp,\vq)$, corresponding to even and odd $N$, respectively. We
will then insert the sum (\ref{Sx-35}) into the corresponding
associativity condition (\ref{Sx-29}) and (\ref{Sx-31}), not in the
original form
\begin{eqnarray*}
\zeta(\vp,\vq)+\zeta(\vq,\ver)=\zeta(\vp,\ver)+\zeta(\vp-\ver,\vq-\ver),
\end{eqnarray*}
where $\zeta(\vp,\vq)$ stands generically for $\omega(\vp,\vq)$ and
$\xi(\vp,\vq)$, but with a change of variable $\vq-\ver\to \vq$,
\begin{eqnarray}\label{Sx-37}
\zeta(\vp,\vq+\ver)+\zeta(\vq+\ver,\ver)=\zeta(\vp,\ver)+\zeta(\vp-\ver,\vq),
\end{eqnarray}
and substitute
\begin{eqnarray}\label{Sx-38}
\zeta(\vp,\vq)=\sum\limits_{\vec{i},\vec{j}}a_{\vec{i},\vec{j}}~{\mathbf{p}}^{{i}}~{\mathbf{q}}^{{j}},
\end{eqnarray}
where we are using two dimensional vector notation
${\mathbf{p}}^{{i}}\equiv p_{1}^{i_{1}}p_{2}^{i_{2}}$,
${\mathbf{q}}^{{j}}\equiv q_{1}^{j_{1}}q_{2}^{j_{2}}$, and the sum
is over $\{i_{1},i_{2},j_{1},j_{2}\}$ with
$i_{1}+i_{2}+j_{1}+j_{2}=N$. Equation (\ref{Sx-37}) then becomes:
\begin{eqnarray}\label{Sx-39}
\sum\limits_{\vec{i},\vec{j},\vec{k}}a_{\vec{i},\vec{j}}~{\mathbf{p}}^{{i}}{\mathbf{q}}^{{j}-{k}}{\mathbf{r}}^{{k}}
\left(\begin{array}{c}
{j}\\
{k}
\end{array}
\right)+
\sum\limits_{\vec{i},\vec{j},\vec{k}}a_{\vec{i},\vec{j}}~{\mathbf{q}}^{{i}-{k}}{\mathbf{r}}^{{j}+{k}}
\left(\begin{array}{c}
{i}\\
{k}
\end{array}
\right)-
\sum\limits_{\vec{i},\vec{j},\vec{k}}a_{\vec{i},\vec{j}}~{\mathbf{p}}^{{i}-{k}}{\mathbf{q}}^{{j}}{\mathbf{r}}^{{k}}(-1)^{{k}}
\left(\begin{array}{c}
{i}\\
{k}\end{array}\right)
=\sum\limits_{\vec{i},\vec{j},\vec{k}}a_{\vec{i},\vec{j}}~{\mathbf{p}}^{{i}}{\mathbf{r}}^{{j}},\nonumber\\
\end{eqnarray}
where $\left(\begin{array}{c}
{i}\\
{k}
\end{array}
\right)\equiv\left(\begin{array}{c}
i_{1}\\
k_{1}
\end{array}
\right) \left(\begin{array}{c}
i_{2}\\
k_{2}
\end{array}
\right)$ etc. and $(-l)^{{k}}\equiv(-1)^{k_{1}+k_{2}} $. It is not
hard to derive the recurrence relation for $a_{\vec{i},\vec{j}}$
from this equation. However, the limits on the indices requires
careful attention. We will not go into this tedious discussion and
simply write down the solution
\begin{eqnarray}\label{Sx-40}
\left(\begin{array}{c}
{i}+{k}\\
{k}\end{array}\right)~a_{\vec{i}+\vec{k},\vec{j}-\vec{k}} =(-1)^{{k}}\left(\begin{array}{c}{j}\\
{k}\end{array}\right)~a_{\vec{i},\vec{j}},\qquad\mbox{with}\qquad
\vec{k}\leq \vec{j}.
\end{eqnarray}
We note that $a_{\vec{i},\vec{j}}$ are antisymmetric
\begin{eqnarray}\label{Sx-41}
a_{\vec{j},\vec{i}}=-a_{\vec{i},\vec{j}},
\end{eqnarray}
by the antisymmetry of $\omega(\vp,\vq)$ and $\xi(\vp,\vq)$,
(\ref{Sx-26}) and (\ref{Sx-32}), respectively. Imposing this
condition on the recurrence relation (\ref{Sx-40}) by letting
$\vec{i}+\vec{k}\to \vec{j}$, $\vec{j}-\vec{k}\to \vec{i}$, we get
\begin{eqnarray}\label{Sx-42}
a_{\vec{j},\vec{i}}=(-1)^{{j}-{i}}a_{\vec{i},\vec{j}},
\end{eqnarray}
which implies
\begin{eqnarray}\label{Sx-43}
(j_{1}+j_{2})-(i_{1}+i_{2})~~\mbox{is odd.}
\end{eqnarray}
But this means that the function has odd parity which is only
satisfied by $\xi(\vp,\vq)$. However, there is an except,
\begin{eqnarray}\label{Sx-44}
\vec{i}=(0,1),~~\vec{j}=(1,0),~\mbox{and}~~\vec{k}=(0,0),
\end{eqnarray}
where the generic function $\zeta(\vp,\vq)$ is even and the
recurrence relation (\ref{Sx-40}) yields only an identity
\begin{eqnarray}\label{Sx-45}
a_{0,1,1,0}=a_{0,1,1,0}.
\end{eqnarray}
This is in fact the single possible solution of
$\omega(\vp,\vq)=\vp\wedge\vq$ from (\ref{Sx-35}) observed in
\cite{lizzi2009}.\footnote{ In \cite{lizzi2009}, the form
(\ref{Sx-36}) was obtained from the equation
$\omega(p,q)=\omega(p-q,p)$, which is readily derived from the
associativity condition. However, this equation can be shown to have
a multitude of solutions of the form
$\omega(p,q)=\sum_{n}c_{n}(p\wedge q)^{n}$ [we are grateful to M.
Alishahiha and H Arfaei for pointing this to us]. One has to use the
associativity condition (\ref{Sx-7}) in its entirety to prove that
only $n=1$ is permissible.} We will now proceed to solve the
recurrence relation (\ref{Sx-40}) and find the most general form for
$\xi(\vp,\vq)$: Eq. (\ref{Sx-40}) is
\begin{eqnarray}\label{Sx-46}
a_{\vec{i}+\vec{k},\vec{j}-\vec{k}}=(-1)^{{k}}\frac{{i}!{j}!}{({i}+{k})!({j}+{k})!}~a_{\vec{i},\vec{j}},\qquad
\vec{k}\leq\vec{j},
\end{eqnarray}
which written in the components is
\begin{eqnarray}\label{Sx-47}
a_{i_{1}+k_{1},i_{2}+k_{2};j_{1}-k_{1},j_{2}-k_{2}}&=&(-1)^{k_{1}+k_{2}}~\frac{i_{1}!i_{2}!j_{1}!j_{2}!}{(i_{1}+k_{1})!(i_{2}+k_{2})!(j_{1}-k_{1})!(j_{2}-k_{2})!}~
a_{i_{1},i_{2};j_{1},j_{2}},
\end{eqnarray}
with $0\leq k_{1}\leq j_{1},0\leq k_{2}\leq j_{2}$ and
$k_{1}+k_{2}<j_{1}+j_{2}$. We can start from
$a_{0,1;j_{1},2n-j_{1}}$, where $N=2n+1$ is the degree of the term
in the expansion of $\xi(\vp,\vq)$, and apply (\ref{Sx-47}) to find
all the terms required by associativity, which results in the
expansion
\begin{eqnarray}\label{Sx-48}
\frac{1}{j_{2}+1}\bigg\{(q_1-p_1)^{j_1}\bigg[q_2^{j_{2+1}}-(q_{2}-p_{2})^{j_{2}+1}\bigg]-p_{1}^{j_{1}}p_{2}^{j_{2}+1}\bigg\}
\end{eqnarray}
with $j_{1}+j_{2}=2n$. We have taken care of the intricacies of the
limits in the summations over $k_{1}$ and $k_{2}$. There is a
further subtlety related to the generation of various terms and
their antisymmetric partner terms in the expansion (\ref{Sx-48}) in
the application of (\ref{Sx-47}). The point is that for each $k_{1}$
and $k_{2}$ by applying (\ref{Sx-47}), there is another term
$(k_{1}',k_{2}')$ generated likewise which is the antisymmetric
partner of $(k_{1},k_{2})$, provided
\begin{eqnarray}\label{Sx-49}
k_{1}+k_{1}'=j_{1},\qquad k_{2}+k_{2}'=j_{2}-1.
\end{eqnarray}
However, in the special cases $j_{2}=0$ or $k_{2}=j_{2}$ when
$j_{2}\neq 0$, there are no antisymmetric partners present in
(\ref{Sx-48}), via (\ref{Sx-47}); thus they should be added to
(\ref{Sx-48}) to complete the polynomial of order $N=2n+1$
satisfying both associativity (\ref{Sx-29}) and antisymmetry
(\ref{Sx-32}) and of course unitality (\ref{Sx-11}).  The final
result is
\begin{eqnarray}\label{Sx-51}
\xi_{j_{1}j_{2}}(\vp,\vq)=(q_{1}-p_{1})^{j_{1}}(q_{2}-p_{2})^{j_{2}+1}+p_{1}^{j_{1}}p_{2}^{j_{2}+1}-q_{1}^{j_{1}}q_{2}^{j_{2}+1}.
\end{eqnarray}
where $j_{1}+j_{2}=2n$. Noting that there is no distinction between
direction $1$ and $2$ in the two dimensional space we are
considering , we could start from $a_{1,0;j_{1},j_{2}}$ and arrive
at
\begin{eqnarray}\label{Sx-51-b}
\xi_{j_{1}j_{2}}(\vp,\vq)=(q_{1}-p_{1})^{j_{1}+1}(q_{2}-p_{2})^{j_{2}}+p_{1}^{j_{1}+1}p_{2}^{j_{2}}-q_{1}^{j_{1}+1}q_{2}^{j_{2}},
\end{eqnarray}
which leads to the general solution for $\xi_{N}(\vp,\vq)$ including
all polynomials of order $N$,
\begin{eqnarray}\label{Sx-50}
\xi_{N}(\vp,\vq)=\sum\limits_{n_{1},n_{2}}C_{n_{1}n_{2}}\xi_{n_{1}n_{2}}(\vp,\vq),
\end{eqnarray}
with $n_{1}+n_{2}=N$, and
\begin{eqnarray}\label{Sx-50-c}
\xi_{n_{1}n_{2}}(\vp,\vq)=(q_{1}-p_{1})^{n_{1}}(q_{2}-p_{2})^{n_{2}}+p_{1}^{n_{1}}p_{2}^{n_{2}}-q_{1}^{n_{1}}q_{2}^{n_{2}}.
\end{eqnarray}
We have verified that the expression (\ref{Sx-50-c}) agrees with the
computer generated polynomials $\xi_{3}$, $\xi_{5}$ of order
$N=3,5$. For $N=3$, $\xi_{j_1j_2}$ are given by
\begin{eqnarray}\label{Sx-52}
\xi_{03}(\vp,\vq)&=& 3 p_2^2 q_2 - 3 p_2 q_2^2,\nonumber\\
\xi_{12}(\vp,\vq)&=& p_2^2 q_1 + 2 p_1 p_2 q_2 - 2 p_2 q_1 q_2 - p_1 q_2^2,\nonumber\\
\xi_{21}(\vp,\vq)&=&2 p_1 p_2 q_1 - p_2 q_1^2 + p_1^2 q_2 - 2 p_1
q_1
q_2,\nonumber\\
\xi_{30}(\vp,\vq)&=&3 p_1^2 q_1 - 3 p_1 q_1^2,
\end{eqnarray}
whereas for $N=5$, they read
\begin{eqnarray}\label{Sx-53}
\xi_{05}(\vp,\vq)&=& 5 p_{2}^4 q_{2} - 10 p_{2}^3 q_{2}^2 + 10 p_{2}^2 q_{2}^3 - 5 p_{2} q_{2}^4,\nonumber\\
\xi_{14}(\vp,\vq)&=&p_{2}^4 q_{1} + 4 p_{1} p_{2}^3 q_{2} - 4
p_{2}^3 q_{1} q_{2} - 6 p_{1} p_{2}^2 q_{2}^2 +
 6 p_{2}^2 q_{1} q_{2}^2 + 4 p_{1} p_{2} q_{2}^3 - 4 p_{2} q_{1} q_{2}^3 - p_{1} q_{2}^4, \nonumber\\
\xi_{23}(p,q)&=& 2 p_{1} p_{2}^3 q_{1} - p_{2}^3 q_{1}^2 + 3 p_{1}^2
p_{2}^2 q_{2} - 6 p_{1} p_{2}^2 q_{1} q_{2} +
 3 p_{2}^2 q_{1}^2 q_{2} - 3 p_{1}^2 p_{2} q_{2}^2 + 6 p_{1} p_{2} q_{1} q_{2}^2 - 3 p_{2} q_{1}^2 q_{2}^2 \nonumber\\
 &&+
 p_{1}^2 q_{2}^3 - 2 p_{1} q_{1} q_{2}^3,\nonumber\\
\xi_{32}(\vp,\vq)&=& 3 p_{1}^2 p_{2}^2 q_{1} - 3 p_{1} p_{2}^2
q_{1}^2 + p_{2}^2 q_{1}^3 + 2 p_{1}^3 p_{2} q_{2} -
 6 p_{1}^2 p_{2} q_{1} q_{2} + 6 p_{1} p_{2} q_{1}^2 q_{2} - 2 p_{2} q_{1}^3 q_{2} - p_{1}^3 q_{2}^2 \nonumber\\
 &&+
 3 p_{1}^2 q_{1} q_{2}^2 - 3 p_{1} q_{1}^2 q_{2}^2,\nonumber\\
\xi_{41}(\vp,\vq)&=& 4 p_{1}^3 p_{2} q_{1} - 6 p_{1}^2 p_{2} q_{1}^2
+ 4 p_{1} p_{2} q_{1}^3 - p_{2} q_{1}^4 + p_{1}^4 q_{2} -
 4 p_{1}^3 q_{1} q_{2} + 6 p_{1}^2 q_{1}^2 q_{2} - 4 p_{1} q_{1}^3
 q_{2},\nonumber\\
\xi_{50}(\vp,\vq)&=&5 p_1^4 q_1 - 10 p_1^3 q_1^2 + 10 p_1^2 q_1^3 -
5 p_1 q_1^4.
\end{eqnarray}
Noting that any odd function of a two dimensional vector $\vp$ can
be expanded as
\begin{eqnarray}\label{S56}
\eta_{2}(\vp)\equiv
\sum\limits_{n=1}^{\infty}\sum\limits_{\ell=0}^{2n+1}C_{\ell,2n+1-\ell}~p_{1}^{\ell}p_{2}^{2n+1-\ell},
\end{eqnarray}
from (\ref{Sx-50-c}), we observe that in general, $\xi(\vp,\vq)$ is
given by
\begin{eqnarray}\label{S57}
\xi(\vp,\vq)=\eta_{2}(\vq)-\eta_{2}(\vp)+\eta_{2}(\vp-\vq),
\end{eqnarray}
with $\eta_{2}(\vp)$ an arbitrary odd function in the form
(\ref{S56}). Thus we have found the most general form for
$\alpha(\vp,\vq)$, describing the translational invariant
star-product of function of two variables, as
\begin{eqnarray}\label{Sx-54}
\alpha({{\vp}},{{\vq}})=\sigma(\vp,\vq)+i\omega({{\vp}},{{\vq}}),
\end{eqnarray}
with
\begin{eqnarray}\label{Sx-54-a}
\sigma(\vp,\vq)=\eta(\vq)-\eta(\vp)+\eta(\vp-\vq),
\qquad\mbox{where}\qquad \eta(\vp)\equiv
\eta_{1}(\vp)+i\eta_{2}(\vp).
\end{eqnarray}
In (\ref{Sx-54}), $\omega({{\vp}},{{\vq}})=\vp \wedge \vq$, and in
(\ref{Sx-54-a}), $\eta_{1}({{\vp}})$ is an arbitrary even function
of ${{\vp}}$ and $\eta_{2}(\vp)$ an arbitrary odd function $\vp$,
satisfying $\eta_{1}(\vec{0})=\eta_2(\vec{0})=0$, and
$\eta(-\vp)=\eta^{*}(\vp)$.
\par
At the end it is of interest to obtain the form of the
star-commutation of coordinates $x_{1}$ and $x_{2}$, derived for the
above star-algebra product (see (4.25) in \cite{lizzi2009})
\begin{eqnarray*}
[x_{i},x_{j}]_{\star}=i\theta_{ij},
\end{eqnarray*}
as there are no quadratic term in $\xi(\vp,\vq)$, and in the real
part of $\alpha(\vp,\vq)$ only $\eta_{1}(\vp-\vq)$ can contribute
which does not as $\eta_{1}(p)$ is even. In the next section, we
will introduce the noncommutative gauge theory using the general
translational invariant noncommutative star-product
(\ref{Sx-1})-(\ref{Sx-2}). The goal is to study the effect of the
elements of the noncommutative structure function $\alpha(\vp,\vq)$,
i.e. $\eta(\vp)$ and $\omega(\vp,\vq)$ on the divergence properties
of Feynman integrals of this theory.
\section{Translational invariant noncommutative $U(1)$ gauge
theory}\label{action}
\setcounter{equation}{0}\par\noindent Let us start with the
Lagrangian density of translational invariant noncommutative $U(1)$
gauge theory, which is given by the ordinary Lagrangian of QED with
the commutative products replaced by the translational invariant
noncommutative star-product (\ref{Sx-1})-(\ref{Sx-2}). The full
Lagrangian density consists of a gauge/ghost and a fermionic part,
${\cal{L}}={\cal{L}}_{g}+{\cal{L}}_{f}$. The gauge/ghost Lagrangian
is given by
\begin{eqnarray}\label{D1}
{\cal{L}}_{g}=-\frac{1}{4}F_{\mu\nu}\star
F^{\mu\nu}-\frac{1}{2\xi}(\partial_{\mu}A^{\mu})\star(\partial_{\nu}A^{\nu})+\frac{1}{2}
\left(i\bar{c}\star\partial^{\mu}D_{\mu}c-i\partial^{\mu}D_{\mu}c\star\bar{c}\right),
\end{eqnarray}
where the non-Abelian field strength tensor is defined by
\begin{eqnarray}\label{D2}
F_{\mu\nu}(x)=\partial_{\mu}A_{\nu}(x)-\partial_{\nu}A_{\mu}(x)+ig[A_{\mu}(x),A_{\nu}(x)]_{\star}.
\end{eqnarray}
The fermionic part of ${\cal{L}}$ reads
\begin{eqnarray}\label{D3}
{\cal{L}}_{f}=i\bar{\psi}\star\gamma^{\mu}\partial_{\mu}\psi-g\bar{\psi}\star
\gamma^{\mu}A_{\mu}\star\psi-m\bar{\psi}\star\psi.
\end{eqnarray}
It arises from the commutative Dirac Lagrangian ${\cal{L}}_{D}=
\bar{\psi}(x)(i\gamma^{\mu}\partial_{\mu}-m)\psi(x)$ and the minimal
coupling
\begin{eqnarray}\label{D4}
\partial_{\mu}\psi(x)\to D_{\mu}\psi(x)\equiv
\partial_{\mu}\psi(x)+igA_{\mu}(x)\star\psi(x).
\end{eqnarray}
Note that similar to the case of Moyal noncommutativity, the minimal
coupling (\ref{D4}) is not unique. There are two other possibilities
for introducing the gauge fields in the Lagrangian,
\begin{eqnarray}\label{D5}
\partial_{\mu}\psi(x)\to D_{\mu}\psi(x)\equiv
\partial_{\mu}\psi(x)-ig\psi(x)\star A_{\mu}(x).
\end{eqnarray}
and
\begin{eqnarray}\label{D6}
\partial_{\mu}\psi(x)\to D_{\mu}\psi(x)\equiv
\partial_{\mu}\psi(x)+ig[A_{\mu}(x),\psi(x)]_{\star}.
\end{eqnarray}
Whereas in (\ref{D4}) the fermions are in the fundamental
representation and the resulting noncommutative action is invariant
under the transformation
\begin{eqnarray}\label{D7}
\psi(x)\to e^{ig\alpha(x)}\star\psi(x),&\qquad\mbox{and}\qquad&
A_{\mu}(x)\to A_{\mu}(x)-D_{\mu}\alpha(x),
\end{eqnarray}
the fermions in (\ref{D5}) and (\ref{D6}) are in the
anti-fundamental and adjoint representations, respectively, and the
resulting noncommutative actions are invariant under
\begin{eqnarray}\label{D8}
\psi(x)\to \psi(x)\star e^{ig\alpha(x)},&\qquad\mbox{and}\qquad&
A_{\mu}(x)\to A_{\mu}(x)+D_{\mu}\alpha(x),
\end{eqnarray}
and
\begin{eqnarray}\label{D9}
\psi(x)\to e^{ig\alpha(x)}\star\psi(x)\star
e^{-ig\alpha(x)},&\qquad\mbox{and}\qquad& A_{\mu}(x)\to
A_{\mu}(x)-D_{\mu}\alpha,
\end{eqnarray}
respectively. Here $D_{\mu}\alpha(x)\equiv
\partial_{\mu}\alpha-ig[\alpha(x),A_{\mu}(x)]_{\star}$. In this paper, we will work with
fermions in the fundamental representation with ${\cal{L}}_{f}$ from
(\ref{D3}). The Lagrangian density of translational invariant $U(1)$
gauge theory is of course invariant under the global $U(1)$
transformation
\begin{eqnarray}\label{D10}
\delta_{\alpha}\psi(x)=i\alpha\psi(x),\qquad\mbox{and}\qquad
\delta_{\alpha}\bar{\psi}(x)=-i\alpha\bar{\psi}(x).
\end{eqnarray}
Following the standard procedure, the Noether currents corresponding
to the global $U(1)$ transformation can be determined, and it can be
shown that the noncommutative gauge theory described by (\ref{D1})
possesses two different Noether currents\footnote{See
\cite{sadooghi2001} for the arguments leading to the invariant and
covariant currents in Moyal noncommutative $U(1)$ gauge theory.}
\begin{eqnarray}\label{D11}
J_{\mu}(x)=\psi(x)\star\bar{\psi}(x)\gamma_{\mu},\qquad\mbox{and}\qquad
j_{\mu}(x)=\bar{\psi}(x)\gamma_{\mu}\star\psi(x).
\end{eqnarray}
Depending on their transformation properties under local $U(1)$
gauge transformation (\ref{D7}), they will be designated, in the
rest of this article, as covariant and invariant currents,
respectively. Using the equations of motion for $\bar{\psi}(x)$ and
$\psi(x)$
\begin{eqnarray}\label{D12}
\partial_{\mu}\bar{\psi}\gamma^{\mu}=ig\bar{\psi}\gamma^{\mu}\star
A_{\mu}+im\bar{\psi},\qquad\mbox{and}\qquad
\gamma^{\mu}\partial_{\mu}\psi
=-igA_{\mu}\star\gamma^{\mu}\psi-im\psi,
\end{eqnarray}
the classical continuity equations of the invariant and covariant
currents read
\begin{eqnarray}\label{D13}
D_{\mu}J^{\mu}(x)=0,\qquad\mbox{and}\qquad
\partial_{\mu}j^{\mu}(x)=0,
\end{eqnarray}
where the covariant derivative
$D_{\mu}=\partial_{\mu}+ig[A_{\mu},\cdot]_{\star}$. Using further
the trace property of the star-product
\begin{eqnarray}\label{D14}
\int d^{d}x~f(x)\star g(x)=\int d^{d}x~g(x)\star f(x),
\end{eqnarray}
it is easy to check that both currents from (\ref{D11}) lead to the
same conserved charge
\begin{eqnarray}\label{D15}
Q\equiv \int d^{d-1}x j^{0}(x)=\int d^{d-1} x J^{0}(x),
\qquad\mbox{with}\qquad \partial_{0}Q=0.
\end{eqnarray}
Similarly, there are two different axial vector currents
\begin{eqnarray}
J_{\mu,5}(x)&=&\psi(x)\star\bar{\psi}(x)\gamma_{\mu}\gamma_{5},\\
\label{D16}
j_{\mu,5}(x)&=&\bar{\psi}(x)\gamma_{\mu}\gamma_{5}\star\psi(x),
\label{D17}
\end{eqnarray}
arising from the invariance of the Lagrangian density (\ref{D1})
under global $U_{A}(1)$ axial transformation
$\delta_{\alpha}\psi=i\alpha\gamma_{5}\psi$. In the chiral limit,
$m\to 0$, similar classical conservation laws as in (\ref{D13}) hold
also for axial vector currents (\ref{D16}) and (\ref{D17}). We will
compute the quantum corrections (anomalies) to the vacuum
expectation values $
\partial^{\mu}J_{\mu,5}$ and $
D^{\mu}J_{\mu,5}$ in Sec. \ref{anomaly}.
\section{Perturbative dynamics of  translational invariant $U(1)$ gauge theory}\label{perturbative}
\setcounter{equation}{0}\par\noindent In this section, we will first
present the Feynman rules of translational invariant $U(1)$ gauge
theory and determine eventually the Feynman integrals of one-loop
quantum corrections corresponding to fermion and photon propagators
and three-point vertex function. The goal is to clarify the r\^{o}le
played by $\alpha(p,q)$, that characterizes the translational
invariant star-product (\ref{Sx-1})-(\ref{Sx-2}). In particular, we
will show that $\eta(p)$ from (\ref{Sx-54}) does not appear in the
internal loop integrals, and, similar to the case of scalar
$\lambda\varphi^{4}$ theory discussed in \cite{lizzi2009}, the
divergence properties of the Feynman integrals are only affected by
the antisymmetric function $\omega(p,q)$ which is given in
two-dimensional noncommutative space by (\ref{Sx-36}). To start, we
present the Feynman rules corresponding to the translational
invariant $U(1)$ gauge theory described by (\ref{D1}).
\vspace{0.5cm}\par\noindent \underline{\textit{Fermion Propagator:}}
\begin{eqnarray}\label{G1}
\SetScale{0.8}
  \begin{picture}(80,20)(0,0)
  \SetWidth{0.5}
    \Vertex(-60,0){2}
    \ArrowLine(-60,0)(5,0)
    \Vertex(5,0){2}
    \Text(-50,6)[lb]{{\Black{{$\alpha$}}}}
    \Text(0,4)[lb]{{\Black{{$\beta$}}}}
    \Text(-25,8)[lb]{{\Black{{$p$}}}}
    \end{picture}
\hspace{2cm}S_{\alpha\beta}(p)=\left(\frac{i}{\gamma\cdot
p-m}\right)_{\alpha\beta}e^{-2\eta_{1}(p)},
\end{eqnarray}
where $\gamma\cdot p\equiv \gamma_{\mu}p^{\mu}$
\par\vspace{0.3cm}\noindent
\underline{\textit{Photon propagator (in Feynman gauge $\xi=1$):}}
\begin{eqnarray}\label{G2}
\SetScale{0.8}
    \begin{picture}(80,20)(0,0)
    \SetWidth{0.5}
    \Vertex(-130,0){2}
    \Photon(-130,0)(-65,0){3}{6}
    \Vertex(-65,0){2}
    \Text(-78,10)[]{$k$}
    \Text(-105,7)[]{$\mu$}
    \Text(-50,7)[]{$\nu$}
    \end{picture}
\hspace{-0.3cm}
D_{\mu\nu}\left(k\right)=-\frac{ig_{\mu\nu}}{k^{2}}e^{-2\eta_{1}(k)}.
\end{eqnarray}
\par\noindent
\underline{\textit{Ghost propagator:}}
\begin{eqnarray}\label{G3}
\SetScale{0.8}
  \begin{picture}(80,20)(0,0)
  \SetWidth{0.5}
    \Vertex(-130,0){2}
    \Vertex(-65,0){2}
    \DashArrowLine(-130,0)(-65,0){3}
    \Text(-80,8)[lb]{{\Black{{$p$}}}}
    \end{picture}
    \qquad G(p)=\frac{i}{p^{2}}e^{-2\eta_{1}(p)}.
\end{eqnarray}
\par\noindent
\underline{\textit{$\bar{\psi}_{\alpha}A_{\mu}\psi_{\beta}$-Vertex:}}
\vspace{0.5cm}
\begin{eqnarray}\label{G4}
\SetScale{0.6}
  \begin{picture}(80,20)(0,0)
  \SetWidth{0.8}
    \Photon(50,8)(120,8){4}{5}
    \ArrowLine(60,20)(100,20)
    \ArrowLine(0,-40)(50,8)
    \ArrowLine(50,8)(0,40)
    \Text(70,5)[lb]{{\Black{{$A_\mu$}}}}
    \Text(45,15)[lb]{{\Black{{$k$}}}}
    \Text(-15,-27)[lb]{$\bar{\psi}_{\alpha}$}
    \Text(17,-17)[lb]{${p}$}
    \Text(-15,25)[lb]{$\psi_{\beta}$}
    \Text(17,17)[lb]{${q}$}
    \end{picture}
\qquad
V_{\mu,\alpha\beta}(p,q;k)&=&ig(2\pi)^{4}\delta^{4}(p-k-q)(\gamma_{\mu})_{\alpha\beta}e^{\alpha(0,-q)}e^{\alpha(-q,-p)},\nonumber\\
&=&ig(2\pi)^{4}\delta^{4}(p-k-q)(\gamma_{\mu})_{\alpha\beta}e^{[\eta(-p)+\eta(q)+\eta(k)]}~e^{-i\omega(p,q)}.\nonumber\\
\end{eqnarray}
\par\noindent
\underline{\textit{$A_{\mu_{1}}A_{\mu_2}A_{\mu_3}$-Vertex:}}
\vspace{0.5cm}
\begin{eqnarray}\label{G5}
\lefteqn{ \SetScale{0.6}
  \begin{picture}(80,20)(0,0)
  \SetWidth{0.8}
    \Photon(50,8)(120,8){3}{5}
    \Photon(0,-40)(50,8){3}{4}
    \Photon(50,8)(0,40){3}{4}
    \Text(70,8)[lb]{{\Black{{$\mu_{1}$}}}}
    \ArrowLine(60,20)(100,20)
    \Text(43,15)[lb]{{\Black{{$p_1$}}}}
    \Text(-15,-25)[lb]{${\mu_2}$}
    \Text(22,-23)[lb]{${p_2}$}
    \ArrowLine(50,-10)(20,-40)
    \ArrowLine(50,20)(20,+40)
    \Text(-15,25)[lb]{${\mu_3}$}
    \Text(22,21)[lb]{${p_3}$}
    \end{picture}
    \qquad
V_{\mu_1\mu_2\mu_3}(p_1,p_2,p_3)=2g
(2\pi)^{4}\delta^{4}(p_{1}+p_{2}+p_{3})}\nonumber\\
&&\hspace{5cm}\times e^{[\eta(p_{1})+\eta(p_2)+\eta(p_3)]}\sin
\left(\omega(p_{1},p_{2})\right)\nonumber\\
&&\hspace{5cm}\times[g_{\mu_1\mu_2}(p_1-p_2)_{\mu_3}+g_{\mu_1\mu_3}(p_{3}-p_1)_{\mu_2}+g_{\mu_3\mu_2}(p_2-p_3)_{\mu_1}].\nonumber\\
\end{eqnarray}
\newpage
{\underline{\textit{$A_{\mu_{1}}A_{\mu_{2}}A_{\mu_{3}}A_{\mu_{4}}$-Vertex:}}}
\par\vspace{.5cm}\noindent
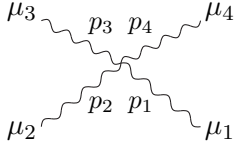
\begin{eqnarray}\label{G6}
\hspace{-4cm}\SetScale{0.5}
  \begin{picture}(80,20)(0,0)
  \SetWidth{0.8}
    \Photon(110,8)(50,40){3}{4}
    \Photon(110,8)(170,40){3}{4}
    \Photon(110,8)(170,-40){3}{4}
    \Photon(110,8)(50,-40){3}{4}
    \Text(87,20)[lb]{$\mu_{4}$}
    \Text(12,20)[lb]{${\mu_3}$}
    \Text(12,-25)[lb]{${\mu_2}$}
    \Text(87,-25)[lb]{${\mu_1}$}
    \Text(58,15)[lb]{{\small{$p_4$}}}
    \Text(58,-15)[lb]{{\small{$p_1$}}}
    \Text(43,-15)[lb]{\small${p_2}$}
    \Text(43,15)[lb]{\small${p_3}$}
    \end{picture}
    \nonumber\\
    \nonumber\\
    \nonumber\\
\lefteqn{\hspace{-3cm}V_{\mu_1\mu_2\mu_3\mu_4}(p_1,p_2,p_3,p_4)=-4ig^{2}(2\pi)^{4}\delta^{4}(p_{1}+p_{2}+p_{3}+p_{4})e^{[\eta(p_1)+\eta(p_2)+\eta(p_3)+\eta(p_4)]}
}\nonumber\\
&&\times\left\{
\sin[\omega(p_{1},p_{2})]\sin[\omega(p_{3},p_{4})]\left(g_{\mu_{1}\mu_{3}}g_{\mu_{2}\mu_{4}}-g_{\mu_{1}\mu_{4}}g_{\mu_{2}\mu_{3}}\right)\right.\nonumber\\
&&\left. +~\sin[\omega(p_{1},p_{3})]
\sin[\omega(p_{2},p_{4})]\left(g_{\mu_{1}\mu_{2}}g_{\mu_{3}\mu_{4}}-g_{\mu_{1}\mu_{4}}g_{\mu_{2}\mu_{3}}\right)\right.\nonumber\\
&&\left. +~
\sin[\omega(p_{1},p_{4})]\sin[\omega(p_{2},p_{3})]\left(g_{\mu_{1}\mu_{2}}g_{\mu_{3}\mu_{4}}-g_{\mu_{1}\mu_{3}}g_{\mu_{2}\mu_{4}}\right)\right\}.
\end{eqnarray}
\par\noindent
{\underline{\textit{$\bar{c}c A_{\mu}$-Vertex:}}}
\begin{eqnarray}\label{G7}
\SetScale{0.5}
  \begin{picture}(80,20) (0,0)
  \SetWidth{0.8}
    \Photon(60,8)(130,8){3}{4}
     \ArrowLine(70,20)(110,20)
    \DashArrowLine(10,-40)(60,8){6}
    \DashArrowLine(60,8)(10,40){6}
    \Text(65,5)[lb]{{\Black{{$A_\mu$}}}}
    \Text(45,15)[lb]{{\Black{{$k$}}}}
    \Text(19,-15)[lb]{${p}$}
    \Text(19,15)[lb]{${q}$}
    \end{picture}
\qquad
G_{\mu}(p,q;k)=2ig(2\pi)^{4}\delta^{4}(p-k-q)p_{\mu}e^{[\eta(-p)+\eta(k)+\eta(q)]}\sin(\omega(p,q)).
\end{eqnarray}
\par\vspace{0.8cm}\par\noindent
According to (\ref{Sx-54}), $\eta(p)=\eta_{1}(p)+i\eta_{2}(p)$ and
$\eta(-p)=\eta^{*}(p)$. Using the above Feynman rules, the one-loop
corrections to fermion and photon propagators and three-point vertex
can be computed.
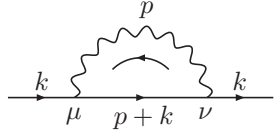
\begin{figure}[htb] \SetScale{0.5}
  \begin{picture}(80,40)(0,0)
  \SetWidth{1.2}
    \ArrowLine(0,0)(50,0)
    \ArrowLine(50,0)(150,0)
    \ArrowLine(150,0)(200,0)
    \PhotonArc(100,0)(50,0,180){4}{9}
    \Text(22,-10)[lb]{\small{\Black{{$\mu$}}}}
    \Text(72,-8)[lb]{\small{\Black{{$\nu$}}}}
    \Text(10,3)[lb]{$k$}
    \Text(85,3)[lb]{$k$}
    \ArrowArc(100,0)(30,45,135)
    \Text(50,30)[lb]{$p$}
    \Text(40,-12)[lb]{\small{$p+k$}}
    \end{picture}
\caption{One-loop fermion self-energy diagram.}
\end{figure}
\par\noindent
The Feynman integral of one-loop fermion-self energy function [Fig.
1] is given by
\begin{eqnarray}\label{G8}
-i\Sigma(k)=-g^{2}\mu^{\epsilon}~e^{2\eta_{1}(k)}\int
\frac{d^{d}p}{(2\pi)^{d}}~
\frac{\gamma_{\mu}[\gamma\cdot\left(k+p\right)+m]\gamma^{\mu}}{p^{2}[(p+k)^{2}-m^2]}.
\end{eqnarray}
\begin{figure}[htb] \SetScale{0.5}
  \begin{picture}(80,40) (0,0)
  \SetWidth{1.2}
    \Photon(-300,0)(-370,0){3}{4}
    \ArrowArc(-270,0)(-30,180,0)
    \ArrowArc(-270,0)(-30,0,180)
    \Photon(-240,0)(-170,0){3}{4}
    \Text(-188,-10)[lb]{\small{\Black{{$\mu$}}}}
    \Text(-90,-8)[lb]{\small{\Black{{$\nu$}}}}
    \Text(-170,8)[lb]{{\Black{{$q$}}}}
    \Text(-110,8)[lb]{{\Black{{$q$}}}}
    \Text(-140,-25)[lb]{{\Black{{$p$}}}}
    \Text(-145,20)[lb]{\small{\Black{{$p+q$}}}}
    \ArrowLine(-230,7)(-190,7)
    \ArrowLine(-350,7)(-310,7)
    \Text(-144,-45)[lb]{(a)}
    \Text(-75,0)[lb]{+}
    \Photon(-105,0)(-35,0){3}{4}
    \PhotonArc(-5,0)(-30,180,0){4}{5}
    \PhotonArc(-5,0)(-30,0,180){4}{5}
    \Photon(25,0)(95,0){3}{4}
    \Text(-55,-10)[lb]{\small{\Black{{$\mu$}}}}
    \Text(43,-8)[lb]{\small{\Black{{$\nu$}}}}
    \Text(-37,8)[lb]{{\Black{{$q$}}}}
    \Text(28,8)[lb]{{\Black{{$q$}}}}
    \Text(-7,-25)[lb]{\small{\Black{{$p$}}}}
    \Text(-15,20)[lb]{\small{\Black{{$p+q$}}}}
    \ArrowLine(-85,7)(-45,7)
    \ArrowLine(45,7)(85,7)
    \ArrowArc(-5,0)(-15,135,-15)
    \Text(-10,-45)[lb]{(b)}
    \Text(58,0)[lb]{+}
    \Photon(160,0)(300,0){3}{6.5}
    \PhotonArc(230,36)(-30,180,0){3}{6}
    \PhotonArc(230,36)(-30,0,180){3}{6}
    \ArrowLine(160,7)(200,7)
    \ArrowLine(260,7)(300,7)
    \ArrowArc(230,36)(-15,135,-15)
    \Text(80,-10)[lb]{\small{$\mu$}}
    \Text(145,-8)[lb]{\small{{$\nu$}}}
    \Text(85,8)[lb]{{\Black{{$q$}}}}
    \Text(140,8)[lb]{{\Black{{$q$}}}}
    \Text(112,39)[lb]{\small{\Black{{$p$}}}}
    \Text(110,-45)[lb]{(c)}
    \Text(165,0)[lb]{+}
    \Photon(380,0)(440,0){3}{4}
    \CArc(470,0)(-30,0,15)
    \CArc(470,0)(-30,25,40)
    \CArc(470,0)(-30,50,65)
    \CArc(470,0)(-30,80,95)
    \CArc(470,0)(-30,105,120)
    \CArc(470,0)(-30,130,145)
    \CArc(470,0)(-30,155,170)
    \CArc(470,0)(-30,180,195)
    \CArc(470,0)(-30,205,220)
    \CArc(470,0)(-30,230,245)
    \CArc(470,0)(-30,255,270)
    \CArc(470,0)(-30,280,295)
    \CArc(470,0)(-30,305,320)
    \CArc(470,0)(-30,330,345)
    \CArc(470,0)(-30,355,360)
    \Photon(500,0)(560,0){3}{4}
    \ArrowLine(390,7)(420,7)
    \ArrowLine(520,7)(550,7)
    \ArrowArc(470,0)(-15,135,-15)
    \Text(190,-10)[lb]{\small{\Black{{$\mu$}}}}
    \Text(275,-8)[lb]{\small{\Black{{$\nu$}}}}
    \Text(205,8)[lb]{{\Black{{$q$}}}}
    \Text(270,8)[lb]{{\Black{{$q$}}}}
    \Text(232,-25)[lb]{\small{\Black{{$p$}}}}
    \Text(225,17)[lb]{\small{\Black{{$p+q$}}}}
    \Text(230,-45)[lb]{(d)}
    \end{picture}
    \vspace{1.5cm}
    \caption{One-loop photon self-energy diagram.}
\end{figure}
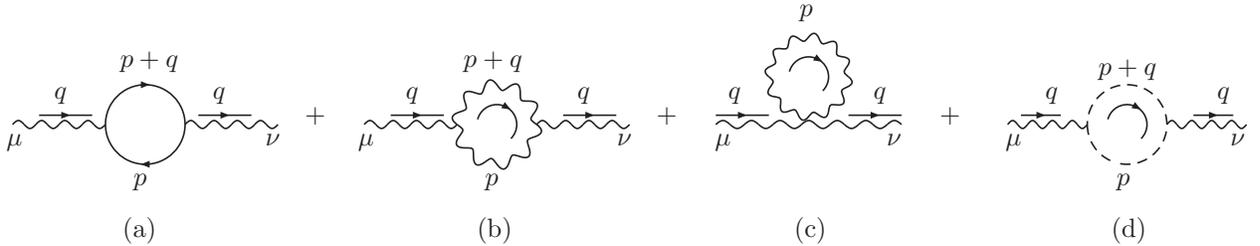
\par\noindent
In Fig. 2, the one-loop diagrams contributing to one-loop photon
self-energy are presented. The corresponding Feynman integrals to
Fig. 2a-2d are given by
\begin{eqnarray}\label{G9}
i\Pi_{\mu\nu}^{(a)}(q)&=&-g^{2}\mu^{\epsilon}~e^{2\eta_{1}(q)}\int
\frac{d^{d}p}{(2\pi)^{d}}~\frac{\mbox{tr}\left((\gamma\cdot
p-m)\gamma_{\mu}(\gamma\cdot(p+q)+m)\gamma_{\nu}\right)}{(p^{2}-m^{2})[(p+q)^{2}-m^{2}]},\nonumber\\
i\Pi_{\mu\nu}^{(b)}(q)&=&\frac{1}{2}(-4g^{2}\mu^{\epsilon})~e^{2\eta_{1}(q)}\int
\frac{d^{d}p}{(2\pi)^{d}}~\frac{\sin^{2}(\omega(p,q))N_{\mu\nu}}{p^{2}(p+q)^{2}},\nonumber\\
i\Pi_{\mu\nu}^{(c)}(q)&=&4g^{2}\mu^{\epsilon}(d-1)g^{\mu\nu}~e^{2\eta_{1}(q)}\int
\frac{d^{d}p}{(2\pi)^{d}}~\frac{\sin^{2}(\omega(p,q))}{p^{2}},\nonumber\\
i\Pi_{\mu\nu}^{(d)}(q)&=&4g^{2}\mu^{\epsilon}~e^{2\eta_{1}(q)}\int
\frac{d^{d}p}{(2\pi)^{d}}~\frac{p_{\mu}(p+q)_{\nu}\sin^{2}(\omega(p,q))}{p^{2}(p+q)^{2}},
\end{eqnarray}
where in $i\Pi_{\mu\nu}^{(b)}(q)$, $N_{\mu\nu}$ is defined by
\begin{eqnarray*}
N_{\mu\nu}&\equiv&\left(g_{\mu\sigma}(-2q-p)_{\rho}+g_{\mu\rho}(q-p)_{\sigma}+g_{\sigma\rho}(2p+q)_{\mu}\right)\nonumber\\
&&\times\left(g^{\sigma\rho}(-2p-q)_{\nu}+\delta^{\sigma}_{~\nu}(2q+p)^{\rho}+\delta_{\nu}^{~\rho}(p-q)^{\sigma}\right).
\end{eqnarray*}
The one-loop diagrams contributing to three-point vertex function
are presented in Fig. 3,
\begin{figure}[htb] \SetScale{0.5}
  \begin{picture}(80,20)(0,0)
    \SetWidth{1.2}
    \ArrowLine(-100,0)(-140,-40)
    \Photon(-140,-40)(-60,-40){3}{4}
    \ArrowLine(-60,-40)(-100,0)
    \Photon(-100,0)(-100,40){3}{3}
    \ArrowLine(-140,-40)(-180,-80)
    \ArrowLine(-20,-80)(-60,-40)
    \ArrowLine(-90,10)(-90,30)
    \ArrowLine(-110,-50)(-90,-50)
    \Text(-40,5)[lb]{\small{\Black{{$q$}}}}
    \Text(-60,5)[lb]{\small{\Black{{$\mu$}}}}
    \Text(-80,-22)[lb]{\small{$\rho$}}
    \Text(-23,-22)[lb]{$\sigma$}
    \Text(-52,-35)[lb]{\small{$p$}}
    \Text(-35,-10)[lb]{\small{$k+p$}}
    \Text(-90,-10)[lb]{\small{$k'+p$}}
    \Text(-5,-40)[lb]{\small{$k$}}
    \Text(-100,-40)[lb]{\small{$k'$}}
    \Text(20,-12)[lb]{$+$}
    \Text(-55,-65)[lb]{(a)}
    \Photon(210,0)(170,-40){3}{4}
    \ArrowLine(250,-40)(170,-40)
    \Photon(250,-40)(210,0){3}{4}
    \Photon(210,0)(210,40){3}{3}
    \ArrowLine(170,-40)(130,-80)
    \ArrowLine(290,-80)(250,-40)
    \ArrowArc(210,-40)(20,45,135)
    \ArrowLine(220,10)(220,30)
    \Text(115,5)[lb]{\small{\Black{{$q$}}}}
    \Text(95,5)[lb]{\small{\Black{{$\mu$}}}}
    \Text(75,-22)[lb]{\small{$\rho$}}
    \Text(132,-22)[lb]{$\sigma$}
    \Text(103,-30)[lb]{\small{$p$}}
    \Text(120,-10)[lb]{\small{$k-p$}}
    \Text(65,-10)[lb]{\small{$k'-p$}}
    \Text(150,-40)[lb]{\small{$k$}}
    \Text(55,-40)[lb]{\small{$k'$}}
     \Text(100,-65)[lb]{(b)}
    \end{picture}
     \vspace{2cm}
    \caption{Diagrams contributing to three-point function.}
\end{figure}
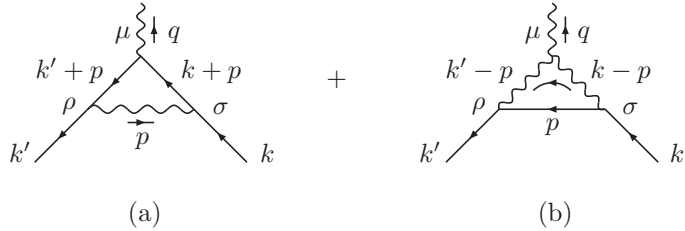
\vspace{0cm}\par\noindent and the corresponding Feynman integrals
are given by
\begin{eqnarray}\label{G10}
V_{\mu}^{(a)}(k,k';q=k-k')&=&g^{3}\mu^{3\epsilon/2}~e^{[\eta(-k)+\eta(k')+\eta(q)]}~e^{-i\omega(k,k')}\nonumber\\
&&\times\int
\frac{d^{d}p}{(2\pi)^{d}}~e^{2i\omega(p,q)}\left(\gamma_{\sigma}\frac{1}{[\gamma\cdot(p+k)-m]}\gamma_{\mu}\frac{1}{[\gamma\cdot(p+k')-m]}\gamma^{\sigma}\frac{1}{k^{2}}\right),
\nonumber\\
V_{\mu}^{(b)}(k,k';q=k-k')&=&g^{3}\mu^{3\epsilon/2}~e^{[\eta(-k)+\eta(k')+\eta(q)]}~e^{-i\omega(k,k')}\nonumber\\
&&\times \int  \frac{d^{d}p}{(2\pi)^{d}}
\left(e^{2i[\omega(p,q)+\omega(k,k')]}-1\right)~\frac{\gamma^{\sigma}(\gamma\cdot p+m)\gamma^{\rho}}{(p^{2}-m^{2})(k-p)^{2}(k'-p)^{2}}\nonumber\\
&&\times
\left(g_{\sigma\rho}(2p-k-k')_{\mu}+g_{\sigma\mu}(2k-k'-p)_{\rho}+g_{\mu\rho}(2k'-k-p)_{\sigma}\right).
\end{eqnarray}
Comparing (\ref{G8})-(\ref{G10}) with the one-loop integrals in
commutative $U(1)$-gauge theory, there are phases depending on
$\eta(p)$ and  $\omega(p,q)$, that arise from the definition of
translational invariant star-product and the form of propagators and
vertices. The appearance of momentum dependent phases in the Feynman
integrals is indeed a characteristic feature for noncommutative
quantum field theory. Here, similar to the ordinary Moyal
noncommutative gauge theory, we will classify the Feynman integrals
into two categories of planar and nonplanar integrals: The planar
integrals involve phases that do not depend on the loop integration
momenta. The phases appearing in the nonplanar integrals, however,
depend on loop momentum and cause the mixing of UV and IR
divergencies in the loop integrations \cite{minwalla1999}. Comparing
the one-loop integrals from (\ref{G8})-(\ref{G10}) with their
counterparts in the ordinary Moyal noncommutative $U(1)$ gauge
theory from e.g. \cite{hayakawa1999, binh2003}, it turns out the
loop integrations from (\ref{G8})-(\ref{G10}) are the same as the
loop integrations of the corresponding diagrams in the Moyal
noncommutative case, and the additional phases involving $\eta(p)$
in (\ref{G8})-(\ref{G10}) are only functions of the momenta of
external legs. We conclude therefore that the UV and IR divergence
properties of the above one-loop integrals are similar to the
divergence properties of the integrals appearing in ordinary Moyal
noncommutative gauge theory.
\par
It is easy to see the cancelation of the phases involving
$\eta(\ell)=\eta_{1}(\ell)+i\eta_{2}(\ell)$ in the loop integrations
over $\ell$. The point is that as can be seen from the expressions
for the vertices (\ref{G4})-(\ref{G7}), each vertex contains a sum
$\sum_{i}\eta_{2}(p_{i})$, with $p_{i}$ the \textit{outgoing}
momenta of the legs of the vertex. Now since in a loop, each
internal line of momentum $\ell_{i}$ from a vertex matches a single
internal line from another vertex's leg with opposite momenta, and
also because $\eta_{2}(\ell_{i})$ is odd under $\ell_{i}\to
-\ell_{i}$, all contributions of $\eta_{2}(\ell_i)$ with internal
loop momenta $\ell_i$ cancel out for all $i$. On the other hand,
although the contributions of even parity $\eta_{1}(\ell_i)$ from
the vertices on both sides of an internal line of momentum
$\ell_{i}$ add to $2\eta_{1}(\ell_{i})$, the total contribution of
$\eta_{1}(\ell_{i})$ cancels due to the presence of an additional
$-2\eta_{1}(\ell_{i})$ from the propagators between two vertices. We
are therefore left only with the antisymmetric $\omega(p,q)$ as a
function of internal loop integration in the nonplanar loop
integrals.
\section{Planar and nonplanar axial anomalies of translational invariant $U(1)$ gauge theory}\label{anomaly}
\setcounter{equation}{0}
\par\noindent
Quantum anomalies of the Moyal noncommutative gauge theory are
widely discussed in the literature \cite{nc-anomaly, sadooghi2000,
sadooghi2001}. In this section, the axial anomalies of the
translational invariant gauge theory corresponding to the covariant
current $J_{\mu,5}$ and the invariant current $j_{\mu,5}$ from
(\ref{D16}) will be determined. We will show, that whereas the axial
anomaly corresponding to the covariant current arises from planar
integrals and is given by a $\star$-modification of the axial
anomaly of commutative $U(1)$ gauge theory, the axial anomaly
corresponding to the invariant current is affected by the above
mentioned UV/IR mixing that arises from the phase factor
$\omega(p,q)$ in the nonplanar Feynman loop integrals. The remaining
phases involving the functions $\eta_{1}(p)$ and $\xi(p,q)$ are
independent of the loop integration momentum and do not affect the
UV and IR behavior of the Feynman integrals. Note that, apart from
the appearance of these additional phase factors, the situation is
similar to the case of Moyal noncommutativity (see e.g. in
\cite{sadooghi2000,sadooghi2001}).
\subsection{Planar axial anomaly}
\noindent As we have noted in the Sec. \ref{review}, the classical
equations of motion of the translational invariant gauge theory
(\ref{D12}) lead, in the chiral limit $m\to 0$, to the classical
continuity equation $D_{\mu}J^{\mu,5}=0$, where
$D_{\mu}=\partial_{\mu}+ig[A_{\mu},\cdot]_{\star}$. It is the
purpose of this section to determine the quantum correction to this
conservation law by computing the vacuum expectation value $\langle
D_{\mu}J^{\mu,5}\rangle$. To do this, let us consider the following
three-point function of one axial vector current and two vector
currents:
\begin{eqnarray}\label{H1}
\Gamma_{P}^{\mu\lambda\nu}(x,y,z)=\langle
T(J_{5}^{\mu}(x)J^{\lambda}(y)J^{\nu}(z))\rangle,
\end{eqnarray}
and determine $\partial_{\mu}^{x}\Gamma_{P}^{\mu\lambda\nu}(x,y,z)$.
The vector currents appearing in (\ref{H1}) are given in
(\ref{D11}). Expressing the currents in terms of fermionic fields
and performing the corresponding Wick contractions, it can be shown
that two triangle diagrams contribute to (\ref{H1}) (see Fig. 4).
\epsfxsize15cm
\begin{figure}[hbt]
\SetScale{0.3}
\begin{center}
\begin{picture} (200,40)(0,20)\SetWidth{0.8}
\Vertex(0,0){5} \Vertex(150,75){5} \Vertex(150,-75){5}
\Line(0,0)(150,75) \Line(0,0)(150,-75) \Line(150,75)(150,-75)
\DashLine(-100,0)(0,0){10} \Photon(150,75)(200,150){3}{4}
\Photon(150,-75)(200,-150){3}{4} \Text(60,0)[]{$\ell$}
\Text(-45,0)[]{$J^{\mu}_{5}\left(x\right)$}
\Text(80,50)[]{$J^{\nu}\left(z\right)$}
\Text(80,-50)[]{$J^{\lambda}\left(y\right)$} \SetScale{0.8}
\LongArrow(65,-10)(65,10) \LongArrow(45,28)(25,18)
\LongArrow(25,-18)(45,-28) \ArrowArc(35,0)(8,90,360) \SetScale{0.3}
\Text(100,0)[]{$+$}
\Vertex(600,0){5} \Vertex(750,75){5} \Vertex(750,-75){5}
\Line(600,0)(750,75) \Line(600,0)(750,-75) \Line(750,75)(750,-75)
\DashLine(540,0)(600,0){10} \Photon(750,75)(800,150){3}{4}
\Photon(750,-75)(800,-150){3}{4}
\Text(240,0)[]{$\ell$} \Text(146,0)[]{$J^{\mu}_{5}\left(x\right)$}
\Text(260,50)[]{$J^{\nu}\left(z\right)$}
\Text(260,-50)[]{$J^{\lambda}\left(y\right)$} \SetScale{0.8}
\LongArrow(290,10)(290,-10) \LongArrow(250,18)(270,28)
\LongArrow(270,-28)(250,-18) \ArrowArcn(260,0)(8,90,180)
\Text(205,-60)[]{(b)} \Text(25,-60)[]{(a)}
\end{picture}
\end{center}
\vskip2.5cm \caption{Triangle diagrams for the anomaly in the axial
vector current $J^{\mu\left(5\right)}\left(x\right)$, indicated by
the dashed line.}
\end{figure}
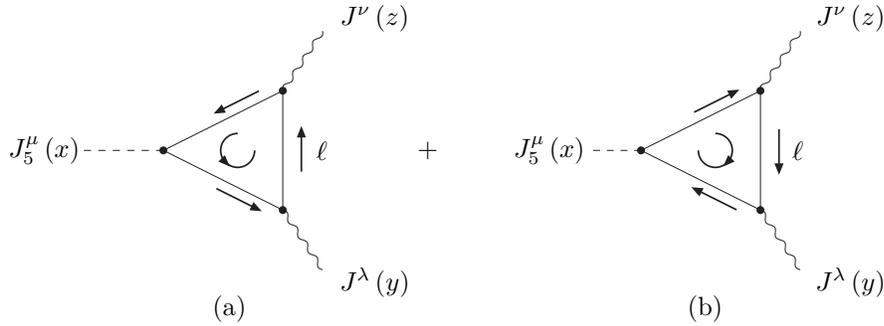
\par\noindent
The corresponding Feynman integrals are given by
\begin{eqnarray}\label{H2}
\lefteqn{\Gamma_{P}^{\mu\lambda\nu}(x,y,z)=
\int\frac{d^{d}k_{2}}{(2\pi)^{d}}\frac{d^{d}k_{3}}{(2\pi)^{D}}e^{-i(k_{2}+k_{3})x}e^{ik_{2}y}
e^{ik_{3}z}
}\nonumber\\
&&\times\int
\frac{d^{d}\ell}{(2\pi)^{D}}\bigg[\mbox{Tr}\left(\gamma^{\mu}\gamma^{5}D^{-1}(\ell+k_{3})
\gamma^{\nu}D^{-1}(\ell)\gamma^{\lambda}D^{-1}(\ell-k_{2})\right)F_{a}(k_{2},k_{3})+\left(
(k_{2},\lambda)\leftrightarrow(k_{3},\nu)\right)F_{b}(k_{2},k_{3})\bigg],\nonumber\\
\end{eqnarray}
where $D(\ell)\equiv \gamma\cdot\ell-m$, and
\begin{eqnarray}\label{H3}
F_{a}(k_{2},k_{3})&=&e^{\alpha(k_{2}+k_{3},\ell+k_{3})+\alpha(-k_{2},\ell-k_{2})+\alpha(-k_{3},\ell)-2[\eta_{1}(\ell-k_{2})+\eta_{1}(\ell)+\eta_{1}(\ell+k_{3})]},\nonumber\\
&=&\exp\left(-[\eta_{1}(k_{2}+k_{3})+\eta_{1}(k_{2})+\eta_{1}(k_{3})]+i[\xi(-k_{2},k_{3})+\omega(k_{2},k_{3})]\right),\nonumber\\
&=&e^{-2[\eta_{1}(k_{2})+\eta_{1}(k_{3})]+\alpha(k_{2}+k_{3},k_{3})},
\end{eqnarray}
arises from the contribution of diagram 4a. The phase factor
appearing on the first line of (\ref{H3}) is simplified using the
definition of $\alpha(p,q)$ in terms of $\eta_{1}(p),\xi(p,q)$ and
$\omega(p,q)$, as well as the properties of $\omega(p,q)$ and
$\xi(p,q)$ [see App. A for a list of these properties]. The
contribution of the diagram 4b is given, as is denoted in the second
term of (\ref{H2}), by replacing $k_{2}\leftrightarrow k_{3}$ as
well as $\lambda\leftrightarrow \nu$. The phase factor corresponding
to diagram 4b, i.e. $F_{b}(k_{2},k_{3})$ can be read from (\ref{H3})
by replacing $k_{2}$ with $k_{3}$ and vice versa. As it turns out,
both phase factors are independent of the loop momentum $\ell$. The
Feynman integral appearing in (\ref{H2}) are therefore planar. The
``planar'' anomaly is given by
$\partial_{\mu}^{x}\Gamma_{P}^{\mu\lambda\nu}$. Taking the partial
derivative with respect to $x^{\mu}$ from
$\Gamma_{P}^{\mu\lambda\nu}$ in (\ref{H2}) and following the same
steps as is described in detail in \cite{sadooghi2000}, we arrive
first at
\begin{eqnarray}\label{H4}
\partial_{\mu}^{x}\Gamma_{P}^{\mu\lambda\nu}(x,y,z)&=&-\frac{i}{4\pi^{2}}\epsilon^{\lambda\nu\alpha\beta}\int\frac{d^{d}k_{2}}{(2\pi)^{d}}\frac{d^{d}k_{3}}{(2\pi)^{d}}
e^{-i(k_{2}+k_{3})x}e^{ik_{2}y}e^{ik_{3}z}e^{-2[\eta_{1}(k_{2})+\eta_{1}(k_{3})]}k_{2\alpha}k_{3\beta}\nonumber\\
&&\times\left(e^{\alpha(k_{2}+k_{3},k_{2})}+e^{\alpha(k_{2}+k_{3},k_{3})}\right).
\end{eqnarray}
Using now the definition of $\langle J^{\mu}_{5}(x)\rangle$ in terms
of the three-point function $\Gamma^{\mu\lambda\nu}$,
\begin{eqnarray}\label{H5} \langle
J^{\mu}_{5}(x)\rangle&=&\frac{1}{2}\int d^{d}y\ d^{d}z\
A_{\lambda}(y)\star \Gamma_{P}^{\mu\lambda\nu}(x,y,z)\star
A_{\nu}(z),
\end{eqnarray}
and the definition of the translational invariant star-product
(\ref{Sx-1})-(\ref{Sx-2}), we get
\begin{eqnarray}\label{H6}
\langle
\partial_{\mu}J^{\mu,5}(x)\rangle=\frac{i}{4\pi^{2}}\epsilon^{\lambda\nu\alpha\beta}\partial_{\alpha}A_{\lambda}(x)\star
\partial_{\beta}A_{\nu}(x).
\end{eqnarray}
After considering the contribution of square and pentagon diagrams
Figs. 5b-5c, we arrive at
\begin{eqnarray}\label{H7}
\langle
D_{\mu}J^{\mu,5}(x)\rangle=\frac{i}{16\pi^{2}}F_{\mu\nu}(x)\star
\tilde{F}^{\mu\nu}(x),
\end{eqnarray}
where $\tilde{F}^{\mu\nu}\equiv
\epsilon^{\mu\nu\rho\sigma}F_{\rho\sigma}$. Thus, similar to the
case of Moyal noncommutativity, the planar (covariant) anomaly
corresponding to the covariant current $J_{\mu,5}(x)$ of
translational invariant $U(1)$ gauge theory is given by a
star-modification of the axial anomaly of commutative $U(1)$ gauge
theory.
\epsfxsize15cm
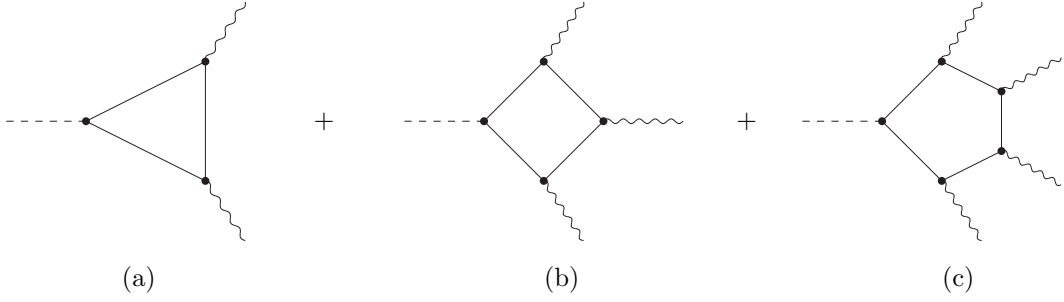
\begin{figure}[htb]
\SetScale{0.3}
\begin{center}
\begin{picture} (350,60)(0,0)\SetWidth{0.8}
\Vertex(0,0){5} \Vertex(150,75){5} \Vertex(150,-75){5}

\Line(0,0)(150,75) \Line(0,0)(150,-75) \Line(150,75)(150,-75)

\DashLine(-100,0)(0,0){10} \Photon(150,75)(200,150){3}{4}
\Photon(150,-75)(200,-150){3}{4}
\Text(20,-60)[]{(a)} \Text(180,-60)[]{(b)} \Text(330,-60)[]{(c)}
\Text(90,0)[]{$+$}
\Vertex(500,0){5} \Vertex(650,0){5} \Vertex(575,75){5}
\Vertex(575,-75){5} \DashLine(400,0)(500,0){10} \Line(500,0)(575,75)
\Line(500,0)(575,-75) \Line(650,0)(575,75) \Line(650,0)(575,-75)
\Photon(575,75)(625,150){3}{5} \Photon(575,-75)(625,-150){3}{5}
\Photon(650,0)(750,0){3}{5}
\Text(250,0)[]{$+$}
\Vertex(1000,0){5}
\Vertex(1075,75){5} \Vertex(1075,-75){5} \Vertex(1150,37.5){5}
\Vertex(1150,-37.5){5} \DashLine(900,0)(1000,0){10}
\Line(1000,0)(1075,75) \Line(1000,0)(1075,-75)
\Line(1150,37.5)(1075,75) \Line(1150,-37.5)(1075,-75)
\Line(1150,37.5)(1150,-37.5) \Photon(1075,75)(1125,150){3}{5}
\Photon(1075,-75)(1125,-150){3}{5} \Photon(1150,37.5)(1225,80){3}{5}
\Photon(1150,-37.5)(1225,-80){3}{5}
\end{picture}
\end{center}
\vskip2.5cm \caption{Triangle, square and pentagon diagrams contributing to
the anomaly in the axial vector current, which is indicated by the dashed
line.}
\end{figure}
\subsection{Nonplanar axial anomaly}
\noindent Let us consider the invariant current $j_{\mu,5}$ from
(\ref{D16}). It satisfies the classical conservation law
$\partial_{\mu}j^{\mu}_{5}=0$. This can be shown using the equations
of motion of the translational invariant $U(1)$ gauge theory in the
chiral limit, (\ref{D12}). The quantum anomaly corresponding to this
current can be computed from the definition of $j_{\mu,5}$ in terms
of the three-point function $\Gamma_{NP}^{\mu\lambda\nu}$
\begin{eqnarray}\label{H8} \langle
j^{\mu}_{5}(x)\rangle&=&\frac{1}{2}\int d^{D}y\ d^{D}z\
A_{\lambda}(y)\star \Gamma_{NP}^{\mu\lambda\nu}(x,y,z)\star
A_{\nu}(z),
\end{eqnarray}
where in contrast to (\ref{H1}),
\begin{eqnarray}\label{H9}
\Gamma_{NP}^{\mu\lambda\nu}(x,y,z)=\langle
T(j_{5}^{\mu}(x)J^{\lambda}(y)J^{\nu}(z))\rangle,
\end{eqnarray}
is a time ordered product of one invariant axial current and two
covariant vector currents. Similar to the previous case,
$\Gamma_{NP}^{\mu\lambda\nu}$ receives contribution from two
triangle diagrams from Fig. 4, where $J^{\mu}_{5}$ is replaced by
$j^{\mu}_{5}$. It is given by
\begin{eqnarray}\label{H10}
\lefteqn{\Gamma_{NP}^{\mu\lambda\nu}(x,y,z)=
\int\frac{d^{d}k_{2}}{(2\pi)^{d}}\frac{d^{d}k_{3}}{(2\pi)^{d}}e^{-i(k_{2}+k_{3})x}e^{ik_{2}y}
e^{ik_{3}z}
}\nonumber\\
&&\times\int
\frac{d^{d}\ell}{(2\pi)^{d}}\bigg[\mbox{Tr}\left(\gamma^{\mu}\gamma^{5}{D^{-1}(\ell+k_{3})}
\gamma^{\nu}{D^{-1}(\ell)}\gamma^{\lambda}{D^{-1}(\ell-k_{2})}\right)F_{a}(\ell;k_{2},k_{3})\nonumber\\
&&\hspace{2cm}+
\left((k_{2},\lambda)\leftrightarrow(k_{3},\nu)\right)F_{b}(\ell;k_{2},k_{3})\bigg],
\end{eqnarray}
where the first (second) term is the contribution from diagram 4a
(4b). The phase factor $F_{a}(\ell,k_{2},k_{3})$ in (\ref{H10}) is
given by
\begin{eqnarray}\label{H11}
F_{a}(\ell;k_{2},k_{3})&=&e^{\alpha(k_{2}+k_{3},k_{2}-\ell)+\alpha(-k_{2},\ell-k_{2})+\alpha(-k_{3},\ell)-
2[\eta_{1}(\ell+k_{3})+\eta_{1}(\ell-k_{2})+\eta_{1}(\ell)]},\nonumber\\
F_{b}(\ell;k_{2},k_{3})&=&e^{\alpha(k_{2}+k_{3},k_{3}-\ell)+\alpha(-k_{3},\ell-k_{3})+\alpha(-k_{2},\ell)-
2[\eta_{1}(\ell+k_{2})+\eta_{1}(\ell-k_{3})+\eta_{1}(\ell)]}.
\end{eqnarray}
After simple algebraic manipulations, where the definition of
$\alpha(p,q)$ in terms of $\eta_{1}(p),\xi(p,q)$ and $\omega(p,q)$,
as well as the properties of $\omega(p,q)$ and $\xi(p,q)$ are
used,\footnote{In App. A, we have summarized useful relations for
$\omega(p,q)$ and $\xi(p,q)$.} it can be shown that
$F_{a/b}(\ell;k_{2},k_{3})$ can be separated into an
$\ell$-independent and an $\ell$-dependent part {\small{
\begin{eqnarray}\label{H12}
F_{a}(\ell;k_{2},k_{3})&=&\exp\left(-[\eta_{1}(k_{2}+k_{3})+\eta_{1}(k_{2})+\eta_{1}(k_{3})]+i\xi(-k_{2},k_{3})
-i\omega(k_{2},k_{3})+2i[\omega(\ell,k_{2})+\omega(\ell,k_{3})]\right),\nonumber\\
F_{b}(\ell;k_{2},k_{3})&=&\exp\left(-[\eta_{1}(k_{2}+k_{3})+\eta_{1}(k_{2})+\eta_{1}(k_{3})]+i\xi(-k_{3},k_{2})
+i\omega(k_{2},k_{3})+2i[\omega(\ell,k_{2})+\omega(\ell,k_{3})]\right).\nonumber\\
\end{eqnarray}
}}\par\noindent This is in contrast to the $\ell$-independent
(planar) phase factor that appears in (\ref{H2}). To add the
contributions of both graphs, we will use the fact that
$\xi(-k_{2},k_{3})=\xi(-k_{3},k_{2})$ from (\ref{A11}) and will
separate the $\ell$-dependent and $\ell$-independent part of
$F_{a/b}(\ell;k_{2},k_{3})$ appropriately. Using further the
definition of $\omega(p,q)$ from (\ref{Sx-36}), and building
$\partial_{\mu}\Gamma_{NP}^{\mu\lambda\nu}$, we arrive at (see also
\cite{sadooghi2001} for notations)
\begin{eqnarray}\label{H13}
\partial_{\mu}^{x}\Gamma_{NP}^{\mu\lambda\nu}=\int\frac{d^{d}k_{2}}{(2\pi)^{d}}\frac{d^{d}k_{2}}{(2\pi)^{d}}
e^{-ik_{1}x}e^{ik_{2}y}e^{ik_{3}z}e^{-[\eta_{1}(k_{1})+\eta_{1}(k_{2})+\eta_{1}(k_{3})]+i\xi(-k_{2},k_{3})}[{\cal{A}}^{\lambda\nu}(k_{2},k_{3})+{\cal{R}}^{\lambda\nu}(k_{2},k_{3})].\nonumber\\
\end{eqnarray}
with $k_{1}\equiv k_{2}+k_{3}$. The anomalous part of
$\partial_{\mu}^{x}\Gamma_{NP}^{\mu\lambda\nu}$ is given by
\begin{eqnarray}\label{H14}
{\cal{A}}^{\lambda\nu}(k_{2},k_{3})&\equiv&-2i\int\frac{d^{d}\ell}{(2\pi)^{d}}\bigg[\mbox{Tr}\left(D^{-1}(\ell-k_{2})\gamma^{5}\ell_{\perp}\cdot\gamma^{\perp}D^{-1}(\ell+k_{3})\gamma^{\lambda}
D^{-1}(\ell)\gamma^{\nu}\right)G_{a}(\ell;k_{2},k_{3})\nonumber\\
&&+\left((k_{2},\lambda)\leftrightarrow
(k_{3},\nu)\right)G_{b}(\ell;k_{2},k_{3})\bigg],
\end{eqnarray}
with $\ell_{\perp}=(\ell_{4},\cdots,\ell_{d-1})$, and\footnote{Here,
we have restricted ourselves to noncommutativity between two space
coordinates $x_{1}$ and $x_{2}$.}
\begin{eqnarray}\label{H15}
G_{a}(\ell;k_{2},k_{3})=e^{-ik_{2}\wedge k_{3}+2i\ell\wedge
(k_{2}+k_{3})},\qquad G_{b}(\ell;k_{2},k_{3})=e^{ik_{2}\wedge
k_{3}+2i\ell\wedge (k_{2}+k_{3})},
\end{eqnarray}
and the rest term of $\partial_{\mu}^{x}\Gamma_{NP}^{\mu\lambda\nu}$
by
\begin{eqnarray}\label{H16}
{\cal{R}}^{\lambda\nu}(k_{2},k_{3})&\equiv&i\int\frac{d^{d}\ell}{(2\pi)^{d}}\bigg[\mbox{Tr}\left(D^{-1}(\ell-k_{2})\gamma^{5}\gamma^{\nu}D^{-1}(\ell)\gamma^{\lambda}+
\gamma^{5}D^{-1}(\ell+k_{3})\gamma^{\nu}D^{-1}(\ell)\gamma^{\lambda}\right)G_{a}(\ell;k_{2},k_{3})\nonumber\\
&&\hspace{2cm}+\left((k_{2},\lambda)\leftrightarrow
(k_{3},\nu)\right)G_{b}(\ell;k_{2},k_{3})\bigg].
\end{eqnarray}
After performing an appropriate shift of the integration variable,
the rest term can be shown to vanish and we are therefore left with
the anomalous part, which is the same as appears also in
\cite{sadooghi2001} for Moyal noncommutative $U(1)$ case. Simple
algebraic manipulations lead to
\begin{eqnarray}\label{H17}
\hspace{-0.5cm}{\cal{A}}^{\lambda\nu}(k_{2},k_{3})=-16\
\varepsilon^{\lambda\nu\alpha\beta}k_{2\alpha}k_{3\beta}\int_{0}^{1}d\beta_{1}\int_{0}^{1-\beta_{1}}d\beta_{2}
\int\frac{d^{d}\ell}{(2\pi)^{d}}\
\frac{\ell_{\perp}^{2}F_{a}\left(\ell+k_{2}\beta_{1}-k_{3}\beta_{2};k_{2},k_{3}\right)}{\left(\ell^{2}+\Delta\right)^{3}},
\end{eqnarray}
with $\Delta\equiv
k_{3}^{2}\beta_{1}(1-\beta_{1})+k_{3}^{2}\beta_{2}(1-\beta_{2})+2k_{2}k_{3}\beta_{1}\beta_{2}$.
Following the same steps as is described in detail in
\cite{sadooghi2001}, ${\cal{A}}^{\lambda\nu}$ is given by
\begin{eqnarray}\label{H18}
{\cal{A}}^{\lambda\nu}(k_{2},k_{3})&=&-\frac{2}{\pi^{2}}\varepsilon^{\lambda\nu\alpha\beta}k_{2\alpha}k_{3\beta}\int_{0}^{1}d\beta_{1}\int_{0}^{1-\beta_{1}}d\beta_{2}
\cos[k_{2}\wedge k_{3}(1-2\beta_{1}-2\beta_{2})]\nonumber\\
&&\times \frac{1}{\ln\Lambda^{2}}\left({\cal{E}}_{1}(k_{1},\Delta;\Lambda_{\mbox{\tiny{eff}}})-\frac{k_{1}\circ k_{1}}{8}~{\cal{E}}_{2}(k_{1},\Delta;\Lambda_{\mbox{\tiny{eff}}})\right),
\end{eqnarray}
where $q\circ q\equiv
-q_{\mu}\theta^{\mu\nu}\theta_{\nu\rho}q^{\rho}$ and
$\frac{1}{\Lambda_{\mbox{\tiny{eff}}}^{2}}\equiv
\frac{1}{\Lambda^{2}}+\frac{k_{1}\circ k_{1}}{4}$. Moreover, we have
used
\begin{eqnarray}\label{H19}
{\cal{E}}_{1}(k_{1},\Delta;\Lambda_{\mbox{\tiny{eff}}})&=&\int_{0}^{\infty}\frac{d\rho}{\rho}\exp\left(-\rho\Delta-\frac{1}{\Lambda_{\mbox{\tiny{eff}}^{2}}\rho}
\right)=2K_{0}\left(2\sqrt{\frac{\Delta}{\Lambda_{\mbox{\tiny{eff}}}^{2}}}\right)\stackrel{\Lambda_{\mbox{\tiny{eff}}}\to
\infty}{\simeq} \ln\frac{\Lambda_{\mbox{\tiny{eff}}}
^{2}}{\Delta},\nonumber\\
{\cal{E}}_{2}(k_{1},\Delta;\Lambda_{\mbox{\tiny{eff}}})&=&\int_{0}^{\infty}\frac{d\rho}{\rho^{2}}\exp\left(-\rho\Delta-\frac{1}
{\Lambda_{\mbox{\tiny{eff}}^{2}}\rho}
\right)=2\sqrt{\Delta\Lambda^{2}_{\mbox{\tiny{eff}}}}~K_{1}\left(2\sqrt{\frac{\Delta}{\Lambda_{\mbox{\tiny{eff}}}^{2}}}\right)\stackrel{\Lambda_{\mbox{\tiny{eff}}}\to
\infty}{\simeq}
\Lambda^{2}_{\mbox{\tiny{eff}}}-\Delta\ln\frac{\Lambda_{\mbox{\tiny{eff}}}
^{2}}{\Delta}.\nonumber\\
\end{eqnarray}
Plugging ${\cal{A}}^{\lambda\nu}$ back in (\ref{H13}), and using
\begin{eqnarray}\label{H20}
\langle\partial_{\mu}j^{\mu}_{5}(x)\rangle&=&\frac{1}{2}\int d^{d}y\
d^{d}z\ A_{\lambda}(y)\star
\partial_{\mu}^{x}\Gamma_{NP}^{\mu\lambda\nu}(x,y,z)\star
A_{\nu}(z),
\end{eqnarray}
we arrive, after performing the integration over $y$ and $z$ and
inserting
${\cal{E}}_{i}(k_{1},\Delta;\theta,\Lambda_{\mbox{\tiny{eff}}}),
i=1,2,$ from (\ref{H19}), at
\begin{eqnarray}\label{H21}
\lefteqn{\langle\partial_{\mu}j^{\mu}_{5}(x)\rangle
=-\frac{1}{\pi}~\varepsilon^{\lambda\nu\alpha\beta}\int\frac{d^{d}k_{2}}{(2\pi)^{d}}k_{2\alpha}\tilde{A}_{\lambda}(k_{2})e^{-ik_{2}x}
\int\frac{d^{d}k_{3}}{(2\pi)^{d}}k_{3\beta}\tilde{A}_{\nu}(k_{3})e^{-ik_{3}x}e^{\rho(k_{2},k_{3})}
}\nonumber\\
&&\times\int_{0}^{1}d\beta_{1}\int_{0}^{1-\beta_{1}}d\beta_{2}\cos[k_{2}\wedge k_{3}(1-2\beta_{1}-2\beta_{2})]\nonumber\\
&&\times\frac{1}{\ln\Lambda^{2}}\Bigg[
\left(\ln\frac{1}{\frac{1}{\Lambda^{2}}+
\frac{(k_{1}\circ k_{1})}{4}
}-\ln\Delta\right)-\frac{(k_{1}\circ k_{1})}{8}\left(
\frac{1}{
\frac{1}{\Lambda^{2}}+
\frac{(k_{1}\circ k_{1})}{4}
}-\Delta\ln\frac{1}{\frac{1}{\Lambda^{2}}+\frac{(k_{1}\circ k_{1})}{4}}+\Delta\ln\Delta\right)\Bigg],\nonumber\\
\end{eqnarray}
where the exponent $\rho(k_{2},k_{3})$ on the first line is defined
by
\begin{eqnarray}\label{H20-1}
\rho(k_{2},k_{3})\equiv -\eta_{1}(k_{2}+k_{3})+i\xi(-k_{2},k_{3}).
\end{eqnarray}
Comparing to the Moyal noncommutative case an additional factor
$e^{\rho(k_{2},k_{3})}$ appears on the first line of (\ref{H21}).
The UV/IR behavior of the remaining expression
 is the same as in the Moyal case. In \cite{sadooghi2001}, we have shown that while the above nonplanar anomaly vanishes in the UV limit, $\frac{k_{1}\circ k_{1}}{4}\gg \frac{1}{\Lambda^{2}}$,
 a finite anomaly arises due to the IR singularity for $\frac{k_{1}\circ k_{1}}{4}\ll \frac{1}{\Lambda^{2}}$. In this limit, all terms proportional to $k_{1}\circ k_{1}$
 in (\ref{H21})  can be neglected, and the finite anomaly arises from the factor
 $\frac{1}{\ln\Lambda^{2}}\ln\frac{1}{\frac{1}{\Lambda^{2}}}\stackrel{\Lambda\to \infty}{\longrightarrow} 1$ on the third line of (\ref{H21}).
 After integrating over $\beta_{1}$ and $\beta_{2}$, it is then given by
 \begin{eqnarray}\label{H22}
\langle\partial_{\mu}j^{\mu}_{5}(x)\rangle=-\frac{1}{2\pi^{2}}\varepsilon^{\lambda\nu\alpha\beta}\int_{\frac{k_{1}\circ
k_{1}}{4}\ll \frac{1}{\Lambda^{2}}}\frac{d^{d}k_{2}}{(2\pi)^{d}}
\frac{d^{d}k_{3}}{(2\pi)^{d}}\partial_{\alpha}\tilde{A}_{\lambda}(k_{2})e^{-ik_{2}x}e^{\rho(k_{2},k_{3})}\frac{\sin(k_{2}\wedge
k_{3})}{k_{2}\wedge k_{3}}~
\partial_{\beta}\tilde{A}_{\nu}(k_{3})e^{-ik_{3}x}.\nonumber\\
 \end{eqnarray}
Defining, similar to the Moyal noncommutative case
\cite{sadooghi2001}, a new generalized star-product
\begin{eqnarray}\label{H23}
f(x)\star' g(x)\equiv \int
\frac{d^{d}p}{(2\pi)^{d}}~\frac{d^{d}q}{(2\pi)^{d}}~\tilde{f}(p)~e^{\rho(p,q)}
\frac{\sin(p\wedge q)}{p\wedge q}\tilde{g}(q)e^{-i(p+q)x},
\end{eqnarray}
with the symmetric function $\rho(p,q)\equiv
-\eta_{1}(p+q)+i\xi(-p,q)$ and the antisymmetric construction
$p\wedge q$ defined in (\ref{Sx-36}), the resulting nonplanar
anomaly, that arises due to the UV/IR mixing is then given by
\begin{eqnarray}\label{H24}
\langle\partial_{\mu}j^{\mu}_{5}(x)\rangle=-\frac{1}{2\pi^{2}}\varepsilon^{\lambda\nu\alpha\beta}\int_{\frac{k_{1}\circ
k_{1}}{4}\ll \frac{1}{\Lambda^{2}}}\frac{d^{d}k_{2}}{(2\pi)^{d}}
\frac{d^{d}k_{3}}{(2\pi)^{d}}~F_{\alpha\lambda}(k_{2})e^{-ik_{2}x}\star'
F_{\beta\nu}(k_{3})e^{-ik_{3}x},
\end{eqnarray}
Here, the contributions from square and pentagon diagrams in Fig. 5
are also added to (\ref{H22}).
\section{Concluding remarks}
\par\noindent
According to its definition (\ref{Sx-1})-(\ref{Sx-2}), the
translational invariant noncommutative star-product is characterized
by a function $\alpha(p,q)$, whose dependence on the momenta $p$ and
$q$ is mainly restricted by the associativity condition on this
product. In the first part of this paper, we have determined the
structure of $\alpha(p,q)$, for a general noncommutative case, in
terms of an arbitrary real even function $\eta_{1}(p)$ and two real
antisymmetric functions $\xi(p,q)$ and $\omega(p,q)$ that appear in
the imaginary part of $\alpha(p,q)$ [see (\ref{Sx-19}) for the real
part and (\ref{Sx-24}) for the imaginary part of $\alpha(p,q)$].
Focusing then on a special two-dimensional noncommutative space, we
have derived the general form of $\xi(p,q)$ and $\omega(p,q)$ from a
recursive relation arising from the associativity. We have shown
that $\omega(p,q)$, as an even antisymmetric function, is given by
$\omega(p,q)=p\wedge q$, and $\xi(p,q)$, as an odd antisymmetric
function, is given in terms of an arbitrary real odd function
$\eta_{2}(p)$ [see (\ref{S57})]. Combining $\xi(p,q)$ from
(\ref{S57}) with the real part of $\alpha(p,q)$ appearing in
(\ref{Sx-19}), we have defined an arbitrary function
$\eta(p)=\eta_{1}(p)+i\eta_{2}(p)$, with $\eta_{1}(p)$ an arbitrary
even and $\eta_{2}(p)$ an arbitrary odd function of $p$, satisfying
$\eta_{1}(0)=\eta_{2}(0)=0$. The characteristic function
$\alpha(p,q)$ is then expressed alternatively in terms of $\eta(p)$
and $\omega(p,q)$, i.e. $\alpha(p,q)=\sigma(p,q)+i\omega(p,q)$. Note
that $\sigma(p,q)=\eta(q)-\eta(p)+\eta(p-q)$, from (\ref{Sx-54-a}),
and $\omega(p,q)=p\wedge q$ are two distinct and unique solutions
for the associativity relation
$\alpha(p,q)+\alpha(q,r)=\alpha(p,r)+\alpha(p-r,q-r)$, where
$\alpha(p,q)$ is a generic function of two dimensional momenta $p$
and $q$. It is interesting to look for the solutions of this
characteristic relation for $d$-dimensional vectors $p$ and $q$ for
higher dimensions.
\par
In the second part of the paper, we have explored the effect of
functions  $\eta(p)$ and $\omega(p,q)$ on the divergence properties
of Feynman integrals appearing in the noncommutative $U(1)$ gauge
theory including the translational invariant star-product. At
one-loop level, it turned out that $\eta(p)$ appears only as a
function of external loop momenta, and only $\omega(p,q)$ is
responsible for the UV/IR mixing that appears also in the ordinary
Moyal noncommutative field theory. Using the algebraic properties of
$\eta(p)$, however, it was shown that $\eta(p)$ cancels out of all
internal loop integrations and appears only as a function of
external momenta. It cannot therefore affect the divergence
properties of the Feynman integrals. The general topological
arguments leading to this simple but remarkable result is described
in the last paragraph of Sec. IV. Our findings confirm the fact
indicated in \cite{galluccio2008}, that the UV behavior of
noncommutative theories is in general described by the canonical
commutation relation between the coordinates (\ref{I1}), which is
unchanged between the translational invariant product and the Moyal
as well as Wick-Voros products considered in \cite{galluccio2008,
lizzi2009}.
\par
Finally, the planar and nonplanar anomalies of the above gauge
theory were also discussed. As it turned out the nonplanar anomaly,
once nonvanishing, is given, in contrast of nonplanar anomaly of
ordinary Moyal noncommutativity, as a function of a new generalized
star-product including the symmetric function
$\rho(p,q)=-\eta_{1}(p+q)+i\xi(-p,q)$ \textit{and} the antisymmetric
combination $\omega(p,q)=p\wedge q$. The planar anomaly, however, is
given, as in the ordinary Moyal noncommutativity, by the
star-modification of the well-known Adler-Bell-Jackiw axial anomaly.
\par
In the case of Moyal product, the noncommutative gauge theory
appears in the decoupling limit of string theory, on a brane, where
$\omega(p,q)$ is related to the background bulk antisymmetric field
$B$. Here, in the general noncommutative gauge theory, we have, in
addition the function $\eta(p)$ appearing as a profile function for
each field in the momentum representation. It is intriguing to
explore the string theoretical origin of this factor.
\section{Acknowledgments}
\par\noindent
We thank M. Alishahiha for bringing Ref. \cite{lizzi2009} to our
attention, and H. Arfaei for useful discussions.
\begin{appendix}
\section{Useful relations for $\omega(p,q)$ and $\xi(p,q)$}
\setcounter{equation}{0}
\par\noindent
The antisymmetric functions $\omega(p,q)$ and $\xi(p,q)$ that appear
in the imaginary part of $\alpha(p,q)$ satisfy the following
relations:
\begin{eqnarray}
\omega(p,p)&=&\omega(0,p)=\omega(p,0)=0,\label{A1}\\
\omega(p,q)&=&-\omega(q,p), \label{A2}\\
\omega(-p,-q)&=&\omega(p,q), \label{A3}\\
\omega(p-q,p)&=&\omega(p,q),\label{A4}\\
\omega(-q,p)&=&\omega(p,q),\label{A5}\\
\omega(p-r,q-r)&=&\omega(p,q)+\omega(q,r)-\omega(p,r),\label{A6}
\end{eqnarray}
as well as
\begin{eqnarray}
\xi(p,p)&=&\xi(0,p)=\xi(p,0)=0,\label{A7}\\
\xi(p,q)&=&-\xi(q,p),\label{A8}\\
\xi(-p,-q)&=&-\xi(p,q),\label{A9}\\
\xi(p-q,p)&=&-\xi(p,q),\label{A10}\\
\xi(-q,p)&=&\xi(-p,q),\label{A11}\\
\xi(p-r,q-r)&=&\xi(p,q)+\xi(q,r)-\xi(p,r).\label{A12}
\end{eqnarray}

\end{appendix}

\end{document}